\documentclass[reprint, amsmath, amssymb, aps, showkeys]{revtex4-2}
\usepackage{graphicx}
\usepackage{dcolumn}
\usepackage{bm}
\usepackage{amsmath}
\usepackage{float}
\usepackage{subfigure}
\usepackage{booktabs}
\usepackage{CJKutf8}
\usepackage{hyperref}
\usepackage{tikz,xcolor}

\hypersetup{
  colorlinks=true,
  linkcolor=blue,
  urlcolor=blue,
  citecolor=blue,
  linktoc=all
}
\definecolor{lime}{HTML}{A6CE39}
\DeclareRobustCommand{\orcidicon}{%
	\begin{tikzpicture}
	\draw[lime, fill=lime] (0,0) 
	circle [radius=0.16] 
	node[white] {{\fontfamily{qag}\selectfont \tiny ID}};
	\draw[white, fill=white] (-0.0625,0.095) 
	circle [radius=0.007];
	\end{tikzpicture}
	\hspace{-2mm}
}

\foreach \x in {A, ..., Z}{%
	\expandafter\xdef\csname orcid\x\endcsname{\noexpand\href{https://orcid.org/\csname orcidauthor\x\endcsname}{\noexpand\orcidicon}}
}

\begin{document}
\begin{CJK*}{UTF8}{gbsn}

\title{Prospects of constraining on the polarizations of gravitational waves from binary black holes using space- and ground-based detectors}

\author{Jie Wu (吴洁)\orcidA{} }
\author{Jin Li (李瑾)\orcidB{} }
\email{cqujinli1983@cqu.edu.cn}

\affiliation{College of Physics, Chongqing University, Chongqing 401331, China}
\affiliation{Department of Physics and Chongqing Key Laboratory for Strongly Coupled Physics, Chongqing University, Chongqing 401331, China}

\begin{abstract}
The theory of general relativity (GR) predicts the existence of gravitational waves (GWs) with two tensor modes, while alternative theories propose up to six polarization modes. 
In this study, we investigate constraints on GW polarization using a model-independent parametrized post-Einsteinian framework and consider both space- and ground-based detectors. 
By evaluating the capabilities and network performance of LISA, Taiji, TianQin, LIGO, Virgo, KAGRA, and the Einstein Telescope (ET), we analyze their respective contributions. 
Among space-based detectors, Taiji provides the most stringent constraints compared with LISA and TianQin.
Regarding ground-based detectors, LIGO excels in vector modes while ET offers comprehensive constraints across all polarization modes. 
In network scenarios, LISA+TJm performs best, and ET surpasses second-generation detector combinations. 
Furthermore, multiband observations effectively mitigate scalar mode degeneracies thereby significantly enhancing the performance of ground-based detectors. 
Ultimately, combined space- and ground-based observations provide robust constraints on GW polarizations that advance tests for deviations from GR. 
Our findings underscore the potential of future GW missions in refining our understanding of gravitational physics through precise measurements of polarizations.
\end{abstract}

\maketitle
\end{CJK*}

\section{Introduction}
Over the past century, general relativity (GR) has undergone rigorous testing through numerous solar system experiments and astrophysical observations, yielding no definitive evidence suggesting deviations from GR~\cite{test_GR1,test_GR2,test_GR3}. 
Despite successfully passing all experimental tests, there are still indications that GR may require extension due to the challenges from dark energy and dark matter in cosmology~\cite{test_GR4,test_GR5}.
Gravitational waves (GWs) provide a new way to test GR, especially in the strong-field regime where GWs originate from compact objects~\cite{ppE_TianQin}. 
One method to test GR is by detecting additional polarizations of GWs. 
In GR, GWs possess only two tensor modes; however, certain alternative gravity theories predict at most six types of polarizations~\cite{GR_polarization}. 
For instance, Brans-Dicke theory predicts an extra scalar mode~\cite{test_GR2,BD_polarization}, while $f(R)$ theory and Horndeski theory propose two additional scalar modes~\cite{fR_polarization1,fR_polarization2,Horndeski_polarization}. 
Einstein-Aether theory predicts five polarization modes~\cite{Einstein-Aether_polarization1,Einstein-Aether_polarization2}, and some tensor-vector-scalar theories encompass all six polarization modes~\cite{TVS_polarization1,TVS_polarization2}.  
Therefore, detecting additional polarizations of GWs can aid in identifying deviations from GR beyond current experimental limits and unraveling the profound nature of gravity.

Since the first detection of GW by LIGO in GW150914~\cite{GW150914}, nearly a hundred GW events generated by compact binary coalescences (CBCs) have been observed~\cite{GWTC1,GWTC2,GWTC3}.
The analysis of these GW data has consistently demonstrated that all observed events are in agreement with GR so far~\cite{GW_test_GR1,GW_test_GR2,GW_test_GR3,GW_test_GR4,GW_test_GR5,GW_test_GR6,GW_test_GR7}.
The start of the fourth observation run (\href{https://observing.docs.ligo.org/plan/}{O4}), is expected to lead to the detection of more GW events, offering additional opportunities for testing GR.
It is crucial to use multiple detectors capable of capturing signals from various directions to effectively detect the polarization of transient GWs and enhance detection capabilities.~\cite{frequency_response}.
Ground-based detectors operating in the hertz frequency band, such as LIGO~\cite{aLIGO}, Virgo~\cite{aVIRGO}, and KAGRA~\cite{KAGRA}, collaborate to form a network for measuring the polarization of GWs.
For continuous GWs emitted by distant CBCs, a detector with directional change capability alone would suffice for accurate measurements~\cite{frequency_response}.
Space-based GW detectors operating in the millihertz frequency band, including LISA~\cite{LISA}, Taiji~\cite{Taiji}, and TianQin~\cite{TianQin}.
They consist of three spacecraft forming a triangular formation, orbiting either the Sun or Earth.
These configurations enable the detection of additional polarizations due to the change in direction~\cite{test_GR_space}.

Theoretically, the capability of a detector to observe additional polarizations of GWs should be assessed, along with verifying the consistency of the detected GW signal with GR predictions.
One approach involves utilizing a waveform model with purely phenomenological parameters to represent potential deviations from GR and applying constraints to these parameters using observed GW data~\cite{ppE_TianQin}.
Yunes and Pretorius introduced the parametrized post-Einstein (ppE) waveform~\cite{ppE1}, derived from the post-Newtonian (PN) approximation, to effectively parametrize the impact of alternative gravity theories on non-GR polarizations.
Most modified gravity theories, such as Brans-Dicke theory, massive gravity, and bimetric theory, can be characterized within the ppE framework~\cite{ppE_Taiji}.
The ppE framework cannot parametrize all possible deviations from GR, but it does offer a method for testing additional polarizations~\cite{ppE2,ppE3,test_GR_PTA}.

At present, the ppE framework is being employed to impose constraints on additional polarizations for testing GR.
Narikawa and Tagoshi explored the potential of advanced ground-based detectors, such as LIGO, Virgo, and KAGRA, in detecting deviations beyond GR using the ppE framework~\cite{ppE_ground}.
Takeda~\textit{et al.} estimated the accuracy of polarization determination for CBCs using the Fisher information matrix (FIM)~\cite{CBC_FIM1,CBC_FIM2}.
Huwyler~\textit{et al.} employed LISA to detect massive black hole binaries (MBHBs) and test GR by representing ppE waveforms with phase corrections only~\cite{ppE_LISA}.
The potential of Taiji and TianQin in testing GR and constraining non-GR parameters using MBHBs has been discussed in Refs.~\cite{ppE_Taiji,ppE_TianQin}.
Nair~\textit{et al.} investigated the synergistic impact of the Einstein Telescope (ET) and preDECIGO on detector networks, demonstrating improved sensitivity within a specific frequency range~\cite{ppE_ET_DECIGO}.
Xie~\textit{et al.} utilized GW signals from nearly ten thousand double white dwarfs observed by the TianQin and LISA networks to detect additional polarization modes~\cite{DWD}.
In Ref.~\cite{ppE_LISA_Taiji}, Wang and Han utilized the LISA-Taiji network to investigate the ppE parameters of deviations from the GR waveform, demonstrating a significant enhancement in the detection of polarization amplitude through combined observations.

Based on our previous work~\cite{my_paper2,my_paper3,my_paper1}, we investigate the constraints from binary black holes (BBHs) for alternative configurations of space- and ground-based detectors on the detection of additional GW polarizations.
The ppE framework is employed for model-independent testing in GR through the numerical calculation of time-domain GW signals that encompass all polarizations.
The space-based detectors LISA, Taiji, and TianQin are employed for the observation of MBHBs, while the ground-based detectors LIGO, Virgo, KAGRA, and ET observe stellar binary black holes(SBBHs). 
These different networks provide constraint results under distinct observational platforms.
The combination of multiband observations is also taken into consideration.
The long-term observation of SBBHs by space-based detectors effectively resolves the response degeneracy in ground-based detector, enhancing their precision.
The FIM is used to present the constraint results for several typical mass BBHs based on luminosity distance.
The potential impacts of multimessenger observations on improving parameter constraints are also discussed.
Through a systematic research approach, we comprehensively analyze the outcomes derived from additional GW polarizations that are constrained by both space- and ground-based detectors, considering multiple perspectives.

This paper is organized as follows.
In Sec.~\ref{sec:GW_signal}, we introduce time-domain GW signals and ppE parameters for measuring additional polarizations within the ppE framework.
In Sec.~\ref{sec:Detectors_and_response}, we evaluate the performance and configurations of space and ground detectors, analyzing their response functions for various modes.
In Sec.~\ref{sec:Methodology}, we explain the typical BBH sources selected in our paper and the method for calculating the signal-to-noise ratio (SNR) and FIM. 
In Sec.~\ref{sec:Constraints_parameters}, we present the constraint results on ppE parameters, including observations from different networks, multiband, and multimessenger observations. 
Finally, we summarize the results of our research in Sec.~\ref{sec:Conclusion}.

\section{Gravitational wave signal}\label{sec:GW_signal}
\begin{figure*}[ht]
    \begin{minipage}{\textwidth}
        \centering
        \includegraphics[width=0.9\textwidth,
        trim=0 0 0 0,clip]{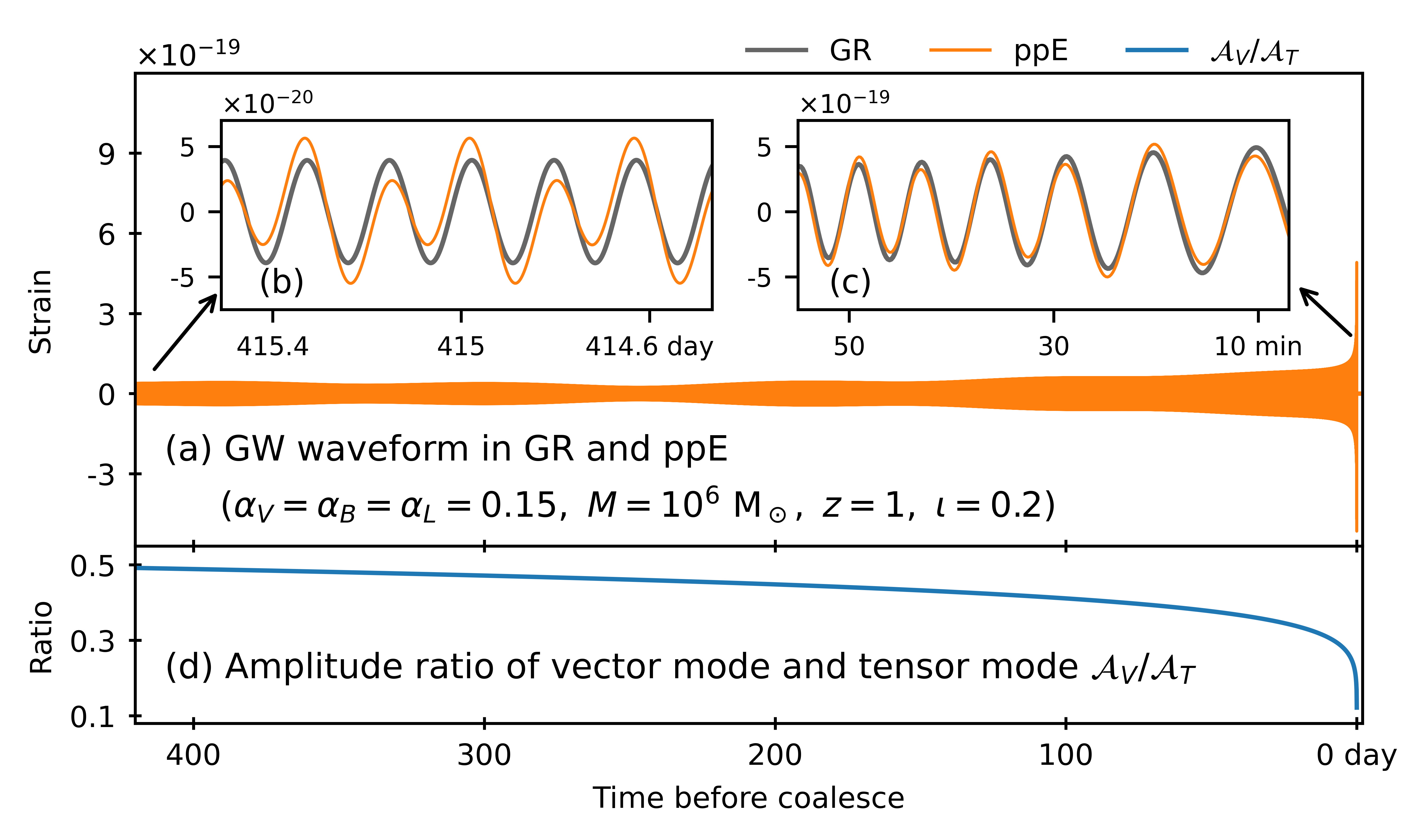}
        \caption{Comparison of time-domain GW waveforms under GR and ppE frameworks. The GW waveform of an MBHB with a total mass of $10^5\ \mathrm{M_\odot }$ is observed by LISA over the year before coalescence. To highlight differences, we set $\alpha_D=0.001$, $\alpha_Q=96/5$, and $\alpha_V=\alpha_B=\alpha_L=0.15$, ensuring $\mathcal{A}_V=\mathcal{A}_B=\mathcal{A}_L$. (a) displays the specific time-domain GW signal. (b) and (c) show the GW waveforms from 415 days and 30 minutes before coalescence, reflecting the early and late inspiral phases, respectively. (d) illustrates the ratio of the amplitude of additional modes to the tensor modes.}\label{fig:signal}
    \end{minipage}
\end{figure*}
In generalized modified gravity theories, GWs can have at most six types of polarizations: two tensor modes ($+$ and $\times$), two vector modes ($X$ and $Y$) and two scalar modes ($B$ and $L$)~\cite{ppE4}.
For detectors, the observed GW strain can be described as a linear combination of different GW polarizations, expressed as
\begin{equation}\label{eq:GW_signal}
    h(t)=\sum_{A}F^A h_A(t) ,
\end{equation}
where $A=\{+,\times,X,Y,B,L\}$ represents for the six polarizations, $h_A(t)$ is the input signal of GWs, and $F^A$ is the angular response function.

The extended ppE framework is utilized to construct a model-independent test for GR, encompassing all GW polarization modes~\cite{ppE2}. 
The amplitude and phase of GWs can be obtained from the metric perturbation and energy evolution, respectively~\cite{ppE_Taiji}.
The process of CBC includes three phases: inspiral, merger, and ringdown. 
The current observations of GW events are transient and do not encompass the early inspiral phase~\cite{GWTC3}.
The successful passing of all current experimental tests in GR provides evidence for the dominance of tensor modes during the merger and ringdown phases.
The contribution of dipole radiation in the early inspiral phase may exceed that during the merger phase~\cite{ppE_Taiji,dipolar}.
The dominance of tensor modes in quadrupole radiation and the dominance of vector and scalar modes in dipole radiation can be reasonably assumed.
Following Refs.~\cite{ppE_PTA,ppE_Taiji,ppE2,ppE4}, the GW waveform under the ppE framework can be written as
\begin{equation}\label{eq:waveform}
    \begin{aligned}
    h_{+} & = \mathcal{A}_T (1+\cos^{2}\iota)/2\cos(2\Phi+2\Phi_0), \\
    h_{\times} & = \mathcal{A}_T \cos\iota\sin(2\Phi+2\Phi_0), \\
    h_{X} & = \mathcal{A}_V \cos\iota\cos(\Phi+\Phi_0), \\
    h_{Y} & = \mathcal{A}_V \sin(\Phi+\Phi_0), \\
    h_{B} & = \mathcal{A}_B \sin\iota\cos(\Phi+\Phi_0), \\
    h_{L} & = \mathcal{A}_L \sin\iota\cos(\Phi+\Phi_0), \\
    \end{aligned}
\end{equation}
with
\begin{equation}\label{eq:amplitude}
    \begin{aligned}
    \mathcal{A}_T&=\frac{4 }{D_L} \left(\frac{G\mathcal{M}}{c^2}\right)^{5/3}\left(\frac{\omega}{c}\right)^{2/3} , \\
    \mathcal{A}_V&= \frac{\alpha_V}{D_L}\left(\frac{G\mathcal{M}}{c^2}\right)^{4/3}\left(\frac{\omega}{c}\right)^{1/3}, \\
    \mathcal{A}_B&= \frac{\alpha_B}{D_L}\left(\frac{G\mathcal{M}}{c^2}\right)^{4/3}\left(\frac{\omega}{c}\right)^{1/3}, \\
    \mathcal{A}_L&= \frac{\alpha_L}{D_L}\left(\frac{G\mathcal{M}}{c^2}\right)^{4/3}\left(\frac{\omega}{c}\right)^{1/3}, \\
    \end{aligned}
\end{equation}
where $\alpha_{V,B,L}$ are the dimensionless ppE parameters, $\mathcal{M}=(m_1m_2)^{3/5}/(m_1+m_2)^{1/5}$ is the chirp mass, $m_1$ and $m_2$ are the masses of BBH, $D_L$ is the luminosity distance, $\iota$ is the inclination angle, $\Phi=\int{\omega}\mathrm{d} t$ is the orbital phase, $\omega$ is the orbital angular frequency, $\Phi_0$ is the initial orbital phase, $G$ and $c$ are the gravitational constant and the speed of light.

The overall evolution of orbital angular frequency can be described by considering the combined effects of dipole and quadrupole radiation~\cite{ppE_PTA,ppE_Taiji}
\begin{equation}\label{eq:angular_frequency}
    \frac{\mathrm{d}\omega}{\mathrm{d}t}=\alpha_D\eta^{2/5}\frac{G\mathcal{M}}{c^3}\omega^3+\alpha_Q\left(\frac{G\mathcal{M}}{c^3}\right)^{5/3}\omega^{11/3} ,
\end{equation}
where $\alpha_D$ and $\alpha_Q$ are the ppE parameters that describe the orbital angular frequency contributions of dipole and quadrupole radiation, $\eta=m_1m_2/M^2$ is the symmetric mass ratio, and $M=m_1+m_2$ is the total mass.

The model-independent GW waveform is described by selecting five dimensionless ppE parameters to constrain the parameters in alternative theories of gravity.
In the case of GR, $\alpha_{D,V,B,L}=0$ and $\alpha_Q=96/5$.
The independence of our analysis is guaranteed by treating these ppE parameters as mutually independent.
Theories often show correlations among these parameters, potentially enhancing the constraints on them~\cite{ppE_Taiji}.
The findings presented here are conservative, and the actual results may be more positive.

The analytical solution for the time evolution function $\omega(t)$ of the orbital angular frequency, as indicated by Eq.~\ref{eq:angular_frequency}), poses a challenging task.
We determine $t(\omega)$ through integration,
\begin{equation}\label{eq:t(w)}
    t=t_0+\int_{\omega(t_0)}^{\omega(t)}{\left(\frac{\mathrm{d}\omega}{\mathrm{d}t}\right)^{-1}} \mathrm{d}\omega,
\end{equation}
which is shown in Refs.~\cite{ppE_PTA,ppE_Taiji}.
The bisection method and other computational techniques are employed to iteratively determine the orbital angular frequency corresponding to a given time point.
By solving point by point, we obtain the value of $\omega(t)$.
Using the above method, we input the calculated $\omega(t)$ into Eqs.~(\ref{eq:GW_signal})$\sim$(\ref{eq:amplitude}) to derive the final GW signal.

In order to better demonstrate the impact of ppE parameters on GW waveforms significantly, we have employed the parameters with large values in Fig.~\ref{fig:signal}.
As shown in Fig.~\ref{fig:signal}(b), a noticeable disparity in amplitude is observed between the ppE waveform and the GR waveform.
The discrepancy arises because the GW frequency resulting from dipole radiation is equivalent to the orbital frequency of the BBH, while quadrupole radiation operates at the twice frequency.
Consequently, this effectively overlays a waveform with a half of the frequency onto the original GR waveform, leading to amplitude modulation at the twice wavelength.
The disparity between the ppE waveform and the GR waveform is significant when the merger is far away but diminishes as the merger approaches, which aligns with our assumption in Eq.~(\ref{eq:waveform}) and current GR testing outcomes (see Figs.~\ref{fig:signal}(b) and (c)). 
The visual representation in Fig.~\ref{fig:signal}(d) shows a gradual decrease in the ratio of additional modes to tensor modes over time, causing the ppE waveform to converge towards the GR waveform.
The amplitude of tensor modes varies with an angular frequency raised to the power of $2/3$, while additional modes vary with an angular frequency raised to the power of $1/3$ according to Eq.~(\ref{eq:amplitude}). 
The amplitude of tensor modes increases more quickly than that of the additional modes, resulting in a declining ratio $\mathcal{A}_V/\mathcal{A}_T \propto \omega^{-1/3}$ as angular frequency increases.

\section{Detectors and response}\label{sec:Detectors_and_response}
\subsection{Space-based detectors}
\begin{figure}[ht]
    \begin{minipage}{\columnwidth}
        \centering
        \includegraphics[width=0.9\textwidth,
        trim=0 0 0 0,clip]{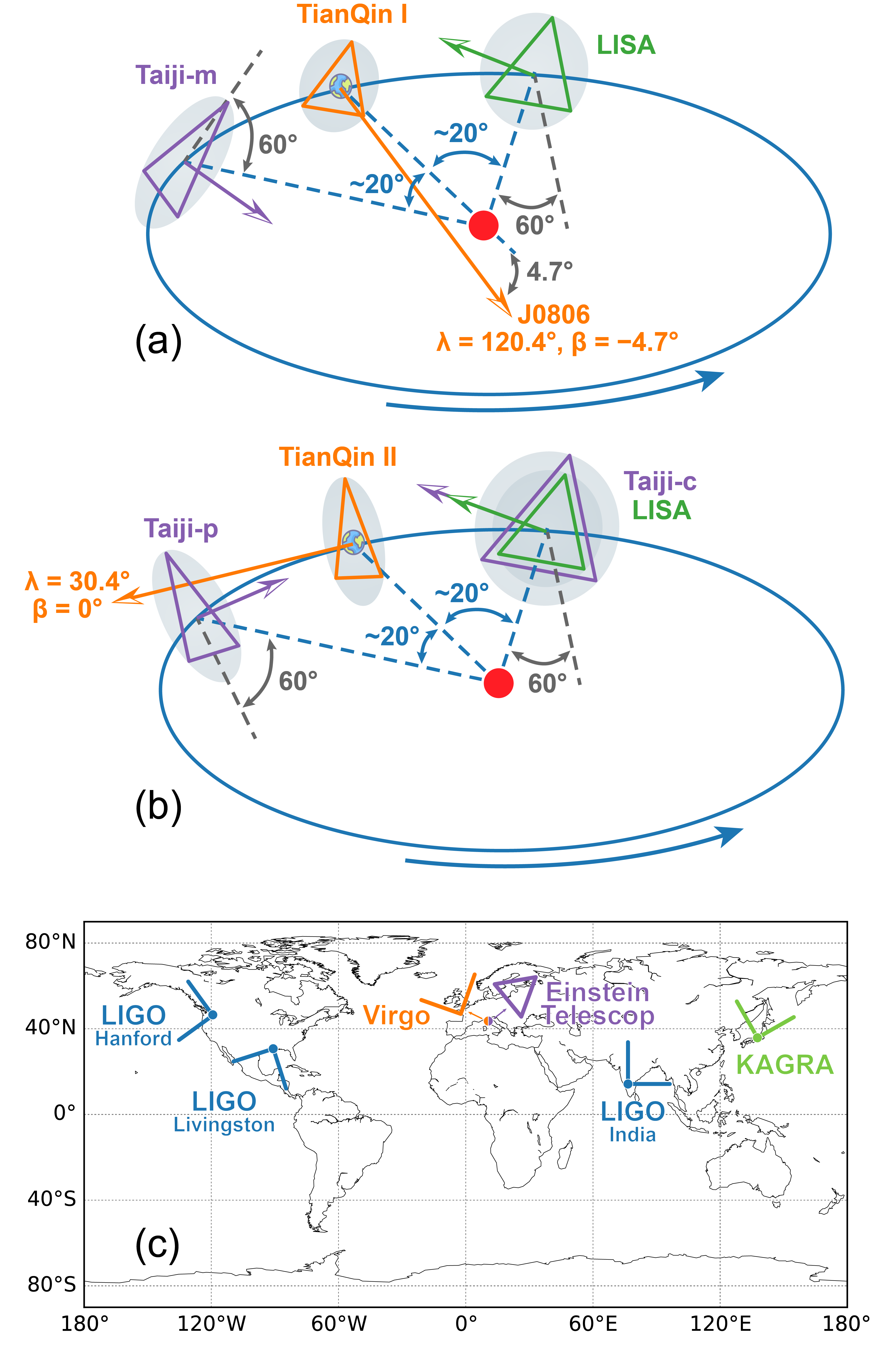}
        \caption{Configuration and location of space- and ground-based detectors (not to scale). (a) and (b) depict alternative orbital configurations of LISA, Taiji-p, Taiji-c, Taiji-m, TianQin I, and TianQin II~\cite{Alternative_LISA_TAIJI1,Alternative_LISA_TAIJI2,Alternative_TianQin,space_network,my_paper2}. The solid arrow on the constellation plane represents the normal direction of the detector constellation plane. (c) shows the geographical locations and azimuth angles of LIGO, Virgo, KAGRA, and ET~\cite{lalsuite1,lalsuite2}.}\label{fig:configuration}
    \end{minipage}
\end{figure}
\begin{figure}[ht]
    \begin{minipage}{\columnwidth}
        \centering
        \includegraphics[width=0.9\textwidth,
        trim=0 0 0 0,clip]{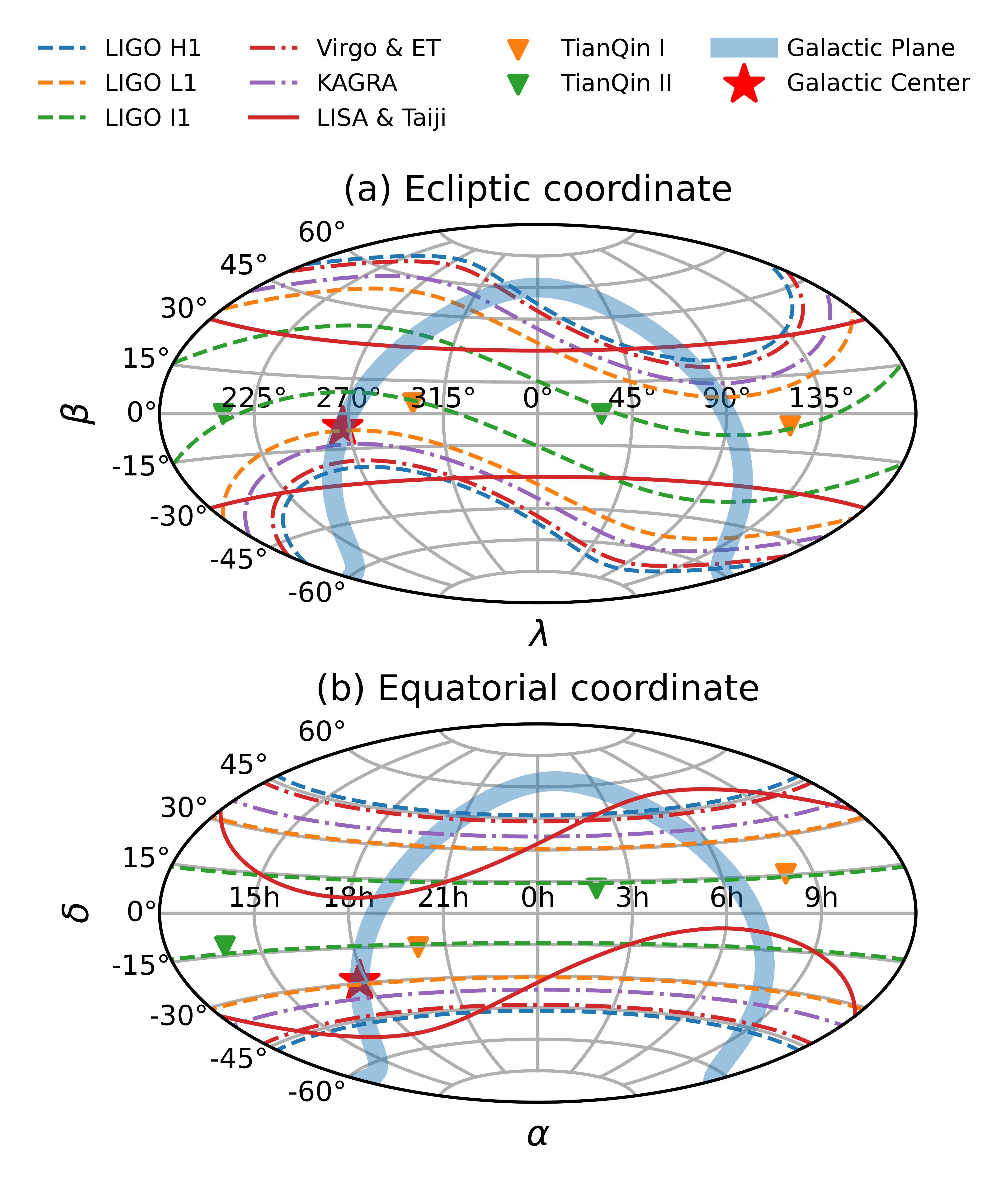}
        \caption{The variation of the normal direction of the detector constellation plane during the observation period in the ecliptic $(\lambda, \beta)$ and equatorial $(\alpha, \delta)$ coordinate systems. This paper uses the obliquity of the ecliptic ($\varepsilon_{2000} = 23^{\circ} 26' 21''$) based on epoch J2000.0 for the conversion between these two coordinate systems~\cite{obliquity_of_ecliptic}.}\label{fig:pointing}
    \end{minipage}
\end{figure}
The LISA, Taiji, and TianQin missions are planned to launch triangular constellations comprising three spacecraft in the 2030s. 
LISA and Taiji will adopt heliocentric orbits, while TianQin will employ geocentric orbits. 
Various alternative orbital configurations are available, as illustrated in Figs.~\ref{fig:configuration} and \ref{fig:pointing}.
The LISA and Taiji missions exhibit a leading or trailing angle of $20^\circ$ with respect to Earth, while maintaining a $60^\circ$ inclination between the constellation plane and the ecliptic plane~\cite{LISA,Taiji}. 
The TianQin constellation plane remains fixed in its normal direction and follows an observation scheme characterized by alternating periods of ``three months on+three months off"~\cite{TianQin}.

The orientation of the constellation plane affects how GW sources appear in the detector coordinate system and influences their response to different polarizations.
The sensitivity of the detectors is directly affected by the length of their arms. 
LISA, Taiji, and TianQin have arm lengths of $2.5\times10^6$ km, $3\times10^6$ km, and $\sqrt{3}\times10^5$ km respectively.
The noise in space-based detectors includes acceleration, displacement, and foreground noise. 
The primary power spectral density (PSD) includes contributions from acceleration and displacement noises, as described in Refs.~\cite{LISA_noise,Taiji_noise,TianQin_noise}. 
The foreground noise comes from galactic binaries in our Milky Way galaxy, creating a peak frequency band of 0.3-3 mHz~\cite{my_paper2}.

\subsection{Ground-based detectors}
Unlike the planned space-based detectors, the operational ground-based detectors primarily include the LVK Collaboration comprising LIGO (4 km), Virgo (3 km), and KAGRA (3 km)~\cite{aLIGO,aVIRGO,KAGRA}.
These detectors are all second-generation with an L-shaped configuration. 
In this paper, we consider LIGO as comprising three operational detectors: LIGO Hanford (H1), LIGO Livingston (L1), and the planned LIGO India (I1). 
The third-generation detector ET is characterized by a triangular shape with three arms, each measuring 10 km~\cite{ET}.

The rotation of the Earth causes ground-based detectors to scan different regions of the sky, as depicted in Fig.~\ref{fig:pointing}.
Due to the obliquity of the ecliptic, the sky regions scanned by space- and ground-based detectors are not parallel.
The performance of ground-based detectors is affected by various types of noise, including quantum noise, seismic noise, gravity-gradient noise, thermal noise, and others. 
In our study, we utilize the design specifications for these detectors and refer to the corresponding PSD in LIGO Document \href{https://dcc.ligo.org/LIGO-T1500293/public}{T1500293} and Refs.~\cite{detector_noise,ET_noise}.

\subsection{Response function}
\begin{figure}[ht]
    \begin{minipage}{\columnwidth}
        \centering
        \includegraphics[width=0.9\textwidth,
        trim=0 0 0 0,clip]{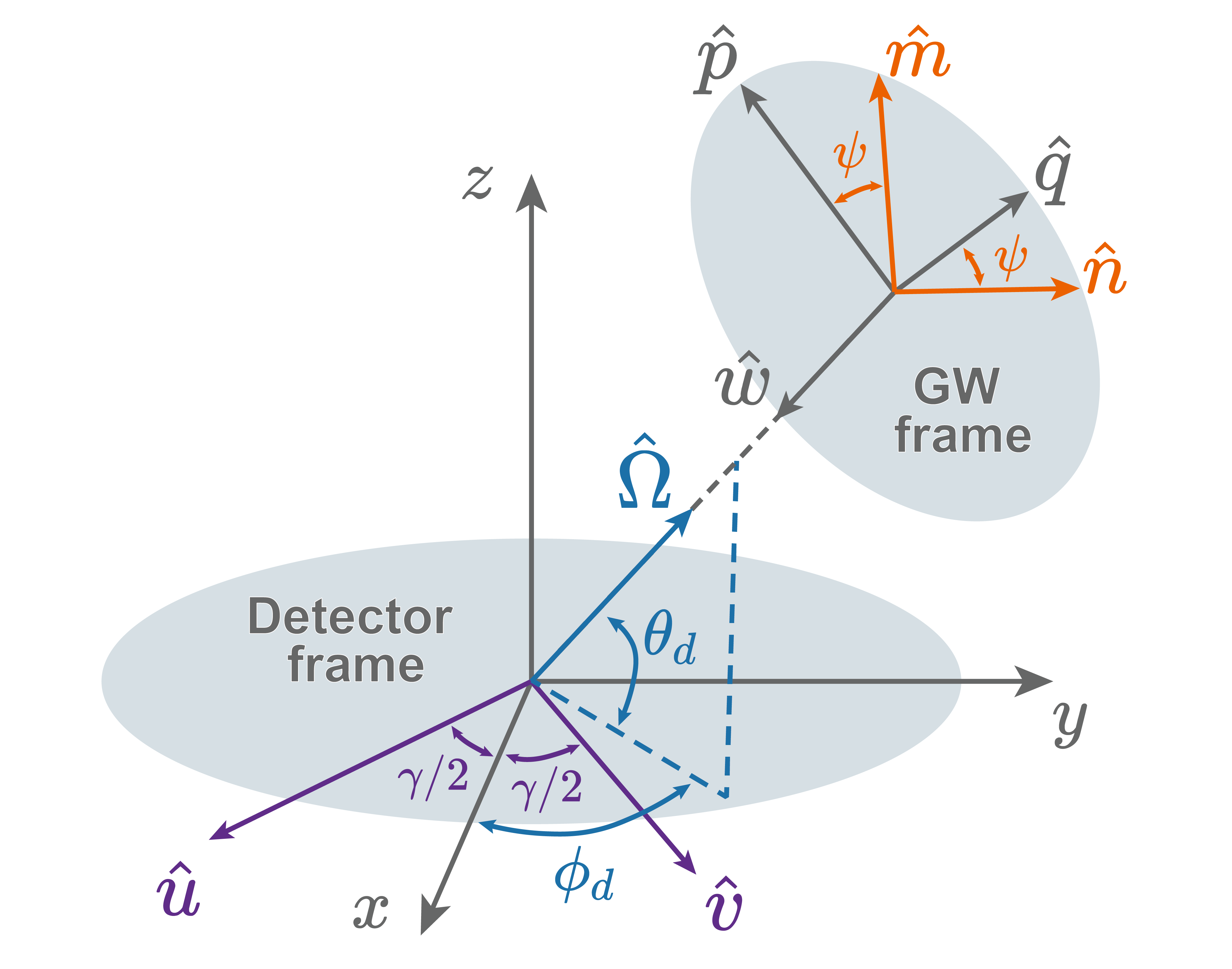}
        \caption{Relationship between the detector coordinate and the GW coordinate~\cite{my_paper3}.}\label{fig:frame}
    \end{minipage}
\end{figure}
\begin{figure*}[ht]
    \begin{minipage}{\textwidth}
        \centering
        \includegraphics[width=0.9\textwidth,
        trim=0 0 0 0,clip]{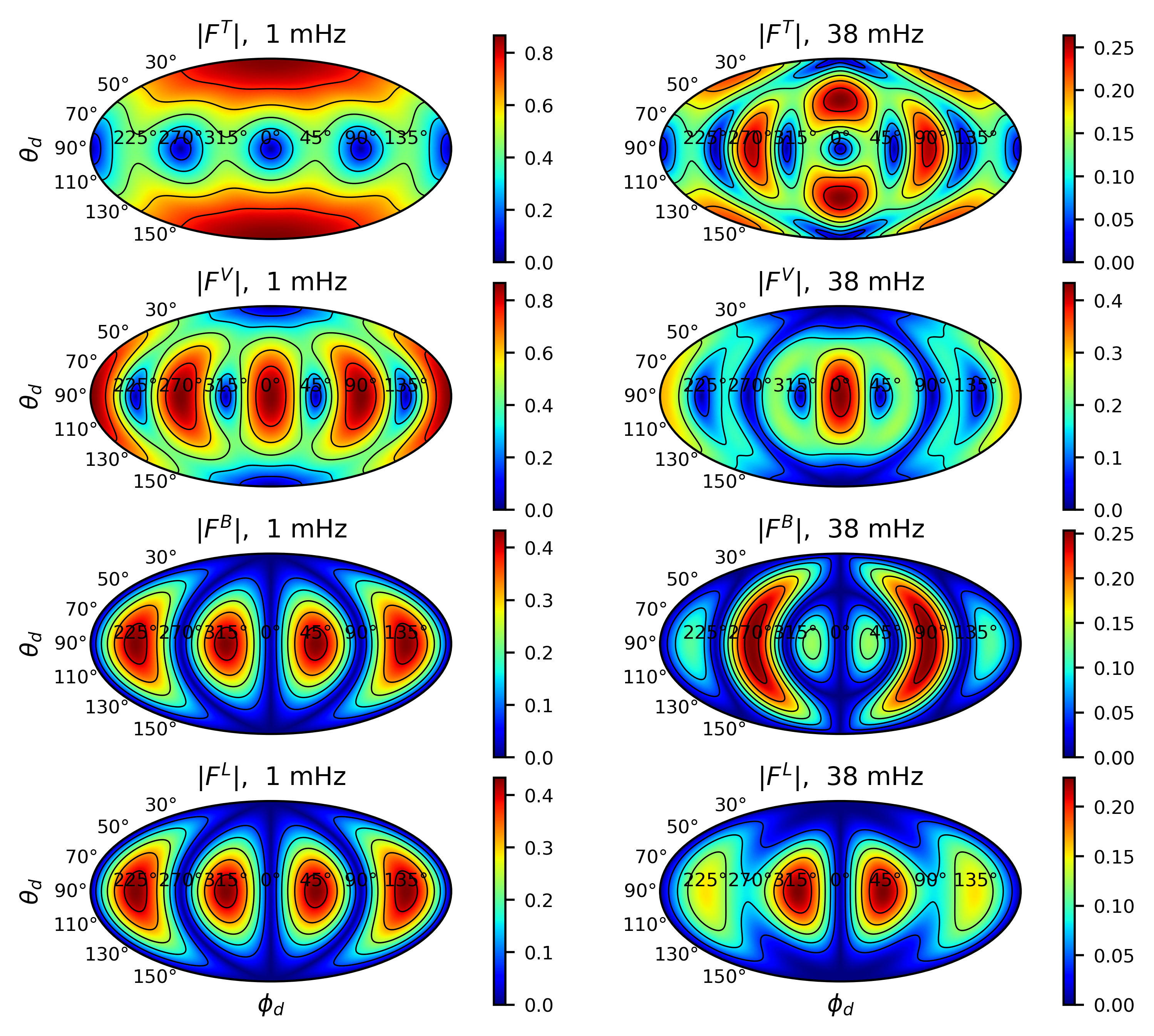}
        \caption{Angular response functions for the combined tensor mode $F^T$, combined vector mode $F^V$, breathing mode $F^B$, and longitudinal mode $F^L$ at LISA detector coordinates $(\phi_d, \theta_d)$. The left panel shows the response functions for GWs at 1 mHz, where the breathing and longitudinal modes exhibit almost indistinguishable characteristics. The right panel displays the response at twice the transfer frequency, $2f_* = 38\ \mathrm{mHz}$, where a distinction between the breathing and longitudinal modes emerges as frequencies increase.}\label{fig:response}
    \end{minipage}
\end{figure*}
The detection of GWs by a dual-arm Michelson interferometer is achieved by measuring the relative change in length between its two arms.
We describe the detector's response using the method outlined in Refs.~\cite{my_paper3,frequency_response}.
The detector coordinates, as illustrated in Fig.~\ref{fig:frame}, are constructed using orthogonal unit vectors \{$\hat{x}, \hat{y}, \hat{z}$\}. 
Similarly, the GW coordinates are constructed using \{$\hat{p}, \hat{q}, \hat{w}$\}. 
Here, $\hat{w}$ represents the propagation direction of the GW, while the unit vector $\hat{\Omega}=-\hat{w}$ denotes the position of the GW source.
The angle $\gamma$ between the two arms is $60^\circ$ in a triangular detector and $90^\circ$ in an L-shaped detector.
For GWs, an additional rotational degree of freedom can be fixed by specifying the polarization angle $\psi$, ultimately using \{$\hat{m}, \hat{n}, \hat{w}$\} to describe the GW.
The angular response function $F^A$ in Eq.~(\ref{eq:GW_signal}) can be expressed as 
\begin{equation}\label{eq:angular_response}
    F^A=D^{ij}e_{ij}^A ,
\end{equation}
where the polarization tensor $e_{ij}^A$ is described using the orthogonal unit vectors mentioned above~\cite{frequency_response,polarization_tensor}:
\begin{equation}
    \begin{aligned}
    e_{ij}^+&=\hat{m}_i\hat{m}_j-\hat{n}_i\hat{n}_j, & e_{ij}^\times&=\hat{m}_i\hat{n}_j+\hat{n}_i\hat{m}_j,\\
    e_{ij}^X&=\hat{m}_i\hat{w}_j+\hat{w}_i\hat{m}_j, & e_{ij}^Y&=\hat{n}_i\hat{w}_j+\hat{w}_i\hat{n}_j,\\
    e_{ij}^B&=\hat{m}_i\hat{m}_j+\hat{n}_i\hat{n}_j, & e_{ij}^L&=\hat{w}_i\hat{w}_j,\\
    \end{aligned}
\end{equation}
and the detector tensor $D^{ij}$ can be written as~\cite{transfer_function}
\begin{equation}
    D^{ij}=\frac{1}{2}[\hat{u}^i\hat{u}^j\mathcal{T}(f,\hat{u}\cdot\hat{w})-\hat{v}^i\hat{v}^j\mathcal{T}(f,\hat{v}\cdot\hat{w})],
\end{equation}
with
\begin{equation}\label{eq:transfer_function}
    \begin{aligned}
    \mathcal{T}(f,\hat{a}\cdot\hat{b})=& \frac{1}{2}\biggr\{\operatorname{sinc}\left[\frac{f}{2f_{*}}(1-\hat{a}\cdot\hat{b})\right] \\
    &\times\exp\left[-i\frac{f}{2f_{*}}\left(3+\hat{a}\cdot\hat{b}\right)\right] \\
    &+\operatorname{sinc}\left[\frac{f}{2f_*}\left(1+\hat{a}\cdot\hat{b}\right)\right] \\
    &\times\exp\biggl[-i\frac{f}{2f_{*}}(1+\hat{a}\cdot\hat{b})\biggr]\biggr\},
    \end{aligned}
\end{equation}
where $\operatorname{sinc}(x)=\sin x/x$, $f_{*}=c/(2\pi L)$ is the transfer frequency and $L$ is the arm length of the detector.
The transfer frequencies for space-based detectors are 19 mHz for LISA, 16 mHz for Taiji, and 275 mHz for TianQin. 
Due to the arm length, the transfer frequency of ground-based detectors exceeds their sensitivity frequency band significantly.

The angular response functions $F^A(\lambda, \beta, \psi)$ or $F^A(\alpha, \delta, \psi)$ are derived by substituting the detector and ecliptic/equatorial coordinates into Eqs.~(\ref{eq:angular_response})$\sim$(\ref{eq:transfer_function}) using Euler rotation conversion.
To present the angular response function concisely, we introduce the combined tensor and vector modes:
\begin{equation}
    \begin{aligned}
    F^T=\sqrt{\left| F^+ \right|^2+\left| F^\times  \right|^2 } ,\\
    F^V=\sqrt{\left| F^X \right|^2+\left| F^Y \right|^2 }.
    \end{aligned}
\end{equation}
The angular response functions of the combined tensor mode $F^T$, combined vector mode $F^V$, breathing mode $F^B$, and longitudinal mode $F^L$ are independent of polarization angle $\psi$~\cite{frequency_response}.
The response of different modes is studied using these four independent modes of $\psi$ and the six basic modes for subsequent research and calculations.
On this basis, we calculate the angular response functions for two different frequencies in the LISA detector coordinates to generate Fig.~\ref{fig:response}.

According to Eq.~(\ref{eq:transfer_function}), in the low-frequency regime $f\ll f_*$, the limit $\mathcal{T} \to 1$ implies $F^B=-F^L$~\cite{frequency_response}. 
The sensitivity frequency bands of ground-based detectors are below this low-frequency limit, making it impossible to distinguish between these two modes.
The potential of space-based detectors to surpass the low-frequency limit and resolve this degeneracy is significant.
The angular response functions for the breathing mode and longitudinal mode are degenerate at every position in Fig.~\ref{fig:response}, indicating their degeneracy at the low-frequency limit.
Beyond the transfer frequency, the optimal response position of the breathing mode shifts from $\phi_d=45^\circ,135^\circ,225^\circ,315^\circ$ to $\phi_d=90^\circ,270^\circ$. 
The shift, which is distinct from the optimal response position of the longitudinal mode, allows for the differentiation between these two modes.

As shown in Fig.~\ref{fig:response}, the optimal response positions of the combined tensor mode and the combined vector mode change with frequency.
The optimal position for the combined tensor mode shifts from a direction perpendicular to the constellation plane to a direction closer to it.
For the combined vector mode, the optimal positions at $\phi_d=90^\circ$ and $\phi_d=270^\circ$ disappear as frequency increases.
In summary, at frequencies beyond the low-frequency limit, there is no degeneracy among the modes, and the optimal response positions generally are not overlapped.

The response at two specific frequencies is illustrated in Fig.~\ref{fig:response}. 
In order to further investigate the relationship between response and frequency, we introduce the angular response function averaged over the source locations~\cite{frequency_response},
\begin{equation}\label{eq:averaged_response}
    R_{A}(f)=\frac{1}{4\pi}\int_{0}^{2\pi}\int_{0}^{\pi}|F^{A}|^{2}\sin\theta_{d}\mathrm{d}\theta_{d}\mathrm{d}\phi_{d},
\end{equation}
where $A=\{T,V,B,L\}$ denotes four modes independent of $\psi$. 
The detector's sensitivity to different frequency modes can be measured by introducing effective strain noise,
\begin{equation}\label{eq:h_eff}
    h^A_{eff}(f)=\sqrt{\frac{S_n(f)}{R_A(f)} },
\end{equation}
where $S_n(f)$ is the noise PSD of the detector.

\begin{figure}[ht]
    \begin{minipage}{\columnwidth}
        \centering
        \includegraphics[width=0.95\textwidth,
        trim=0 0 0 0,clip]{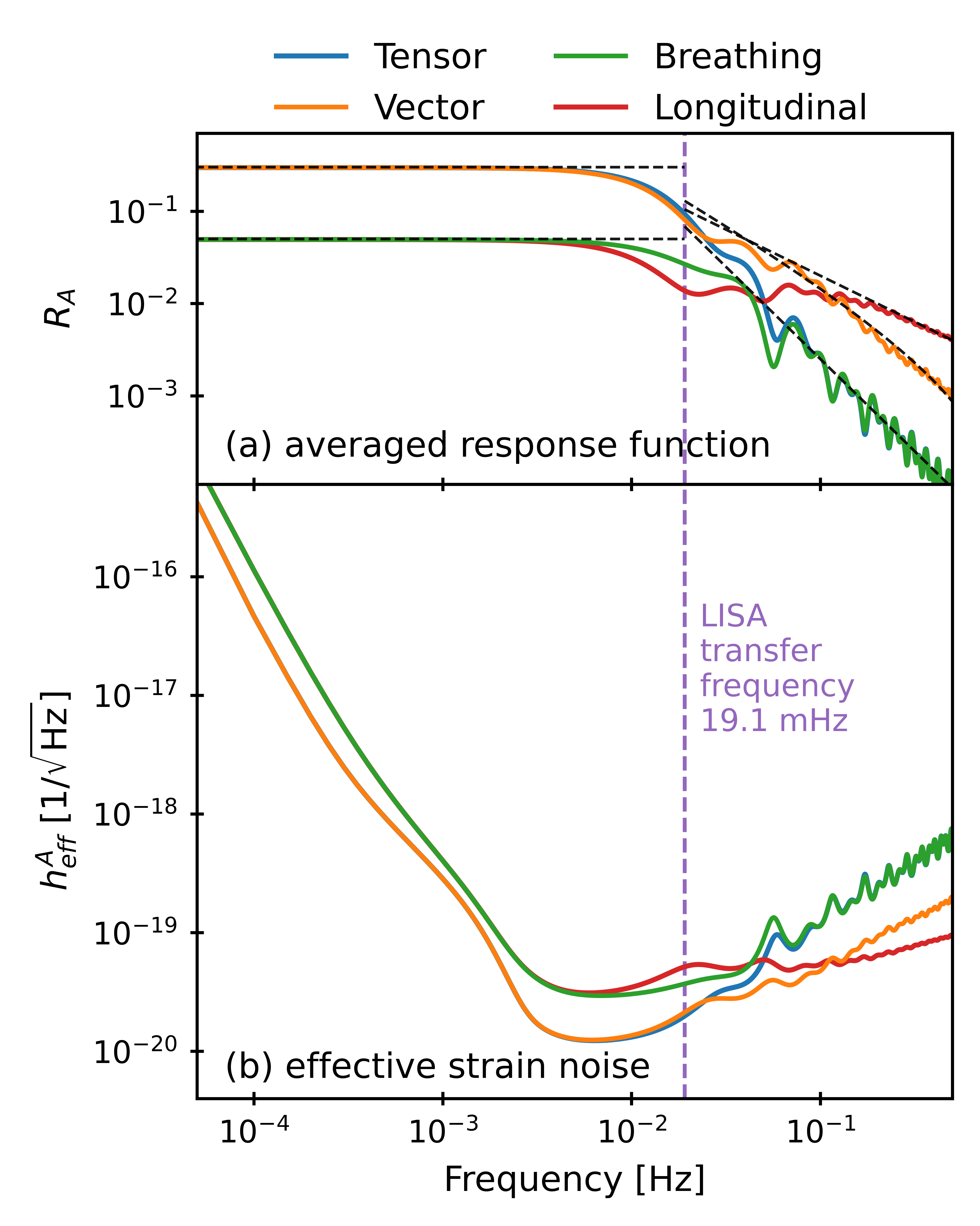}
        \caption{The averaged response functions $R_{A}(f)$ and effective strain noise $h^A_{eff}(f)$ of LISA's four modes vary with frequency. We calculate $R_{A}(f)$ and $h^A_{eff}(f)$ for each frequency point through integration. The purple vertical-dashed line represents LISA's transfer frequency $f_* = 19.1\ \mathrm{mHz}$, and the black dashed line is a fitting approximation of $R_{A}(f)$ (cf. Ref.~\cite{frequency_response}).}\label{fig:sensitivity}
    \end{minipage}
\end{figure}

Through the calculations, we obtain the averaged response functions and effective strain noise for LISA's four modes, as shown in Fig.~\ref{fig:sensitivity}.
At the low-frequency limit, the averaged response function remains constant, with $R_T=R_V=2(\sin^2\gamma)/5$ and $R_B=R_L=(\sin^2\gamma)/15$.
The decrease in $R_A$ becomes evident as the transfer frequency increases, exhibiting three distinct damping trends.
$R_T$ and $R_V$ diverge as the frequency rises, similar to the behavior of $R_B$ and $R_L$.
The differences between $R_T$ and $R_B$ become apparent at low frequencies, but as the frequency increases, these two modes tend to converge.
The average values of these two modes are identical, but their responses at individual positions differ, preventing degeneracy between $R_T$ and $R_B$.

Based on Ref.~\cite{frequency_response}, we analyze the detector's capability to detect various polarizations using the response function.
The sensitive frequency band of ground-based detectors is below the transfer frequency, making it difficult to distinguish between the breathing mode and longitudinal mode.
The degeneracy at nonlow-frequency limits can be resolved by space-based detectors.
In Sec.~\ref{sec:Constraints_parameters}, we simulate GW signals to assess the capability of ground- and space-based detectors in polarization constraint.

\section{Methodology}\label{sec:Methodology}
\subsection{Data analysis}
\begin{figure*}[ht]
    \begin{minipage}{\textwidth}
        \centering
        \includegraphics[width=0.9\textwidth,
        trim=0 0 0 0,clip]{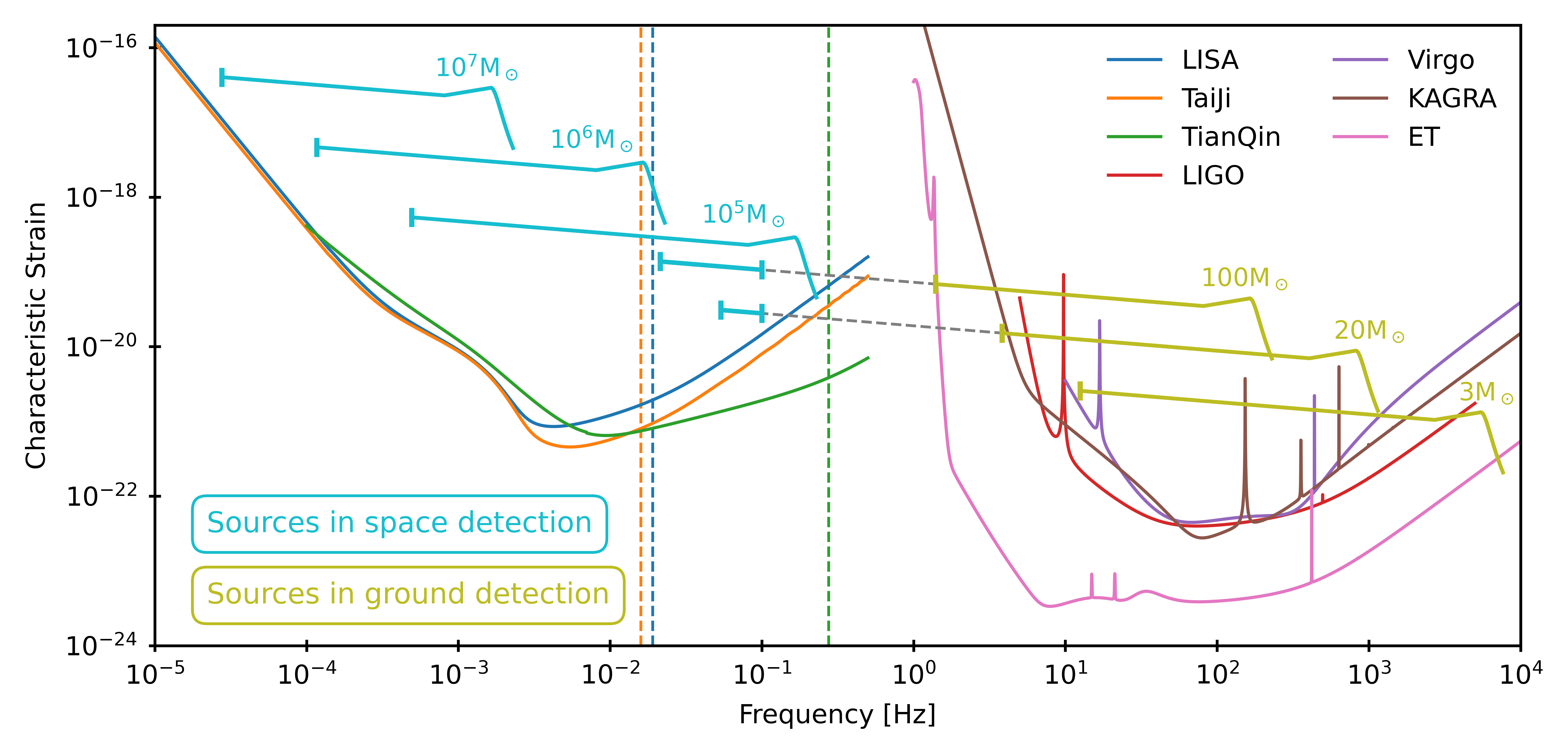}
        \caption{Multiband observation of BBHs with ground- and space-based detectors. The vertical axis represents the dimensionless characteristic strain $\sqrt{fS_n(f)}$. The BBH curve is derive from Ref.~\cite{LISA_noise} and is intended for illustration purposes only, not for calculation purposes. For luminosity distance selection, we set SBBH at $D_L=44.6$ Mpc ($z=0.01$ in $\mathrm{\Lambda CDM} $ cosmological model) and MBHB at $D_L=6.79$ Gpc ($z=1$). The cyan line represents the space detection source, while the olive line represents the ground detection source. The three vertical dashed lines represent the transfer frequencies of the three space-based detectors.}\label{fig:ASD}
    \end{minipage}
\end{figure*}
In general, the SNR $\rho$ of GW can be defined as
\begin{equation}
    \rho^2=(h|h),
\end{equation}
where the inner product $(\cdot|\cdot)$ generalizes the time-domain correlation product and is conventionally defined as
\begin{equation}
    (a|b)=4\text{Re}\left[\int_0^{\infty}\frac{\tilde{a}^*(f)\tilde{b}(f)}{S_n(f)}\mathrm{d}f\right],
\end{equation}
where $\tilde{a}(f)$ and $\tilde{b}(f)$ are the Fourier transforms of $a(t)$ and $b(t)$, respectively.
The total number of GW parameters considered in our study is fourteen, including ppE parameters, which are
\begin{equation}\label{eq:parameters}
    \begin{aligned}
    \boldsymbol{\xi}=\{t_c,m_1,m_2,D_L,\iota,\Phi_0,\phi_e,\theta_e,\\ \psi,\alpha_Q,\alpha_D,\alpha_V,\alpha_B,\alpha_L\},
    \end{aligned}
\end{equation}
where $t_c$ is the coalescence time, $D_L$ is the luminosity distance of the source, $(\phi_e,\theta_e)$ is the sky position, representing ecliptic $(\lambda ,\beta)$ or equatorial $(\alpha ,\delta)$ coordinates.

To evaluate the detectors' performance on different polarizations, we use the FIM:
\begin{equation}\label{eq:FIM}
    \Gamma_{ij}=\left(\frac{\partial h}{\partial\xi_i}\right|\frac{\partial h}{\partial\xi_j}\biggr),
\end{equation}
where $\xi_i$ represents the parameter in Eq.~(\ref{eq:parameters}).
For high SNR, the inverse of the FIM, $\Sigma=\Gamma^{-1}$, is the variance-covariance matrix, with the diagonal elements representing variance~\cite{FIM1,FIM2}.
Thus, the uncertainty $\Delta\xi_i$ of the parameters is given by
\begin{equation}
    \Delta\xi_i=\sqrt{\Sigma_{ii}}.
\end{equation}
When calculating the FIM, we use the numerical differentiation approximation as described in Refs.~\cite{my_paper2,TianQinGBs}.
The total SNR and FIM for the observation network are obtained by summing the inner products calculated by each detector.

\subsection{BBH source selection}\label{sec:source_selection}
The distinct sensitive frequency bands of space- and ground-based detectors facilitate the detection of various GW sources.
Ground-based detectors can observe SBBH's whole CBC in three phases, while space detectors do the same for MBHB.
Additionally, space detectors can also observe the inspiraling SBBH in the low-frequency band. 
The selected BBH sources are shown in Fig.~\ref{fig:ASD}.

The proposed waveform model focuses on the inspiral phase before the binary reaches the innermost stable circular orbit (ISCO). 
Additional polarizations plays a more significant role in this phase than in merger and ringdown phases.
The frequency of ISCO is given by~\cite{ppE_Taiji}
\begin{equation}\label{eq:f_ISCO}
    f_\mathrm{ISCO}=\frac{c^3}{6\sqrt{6}\pi GM }.
\end{equation}
Ground-based detectors typically observe the entry of SBBH GWs into the frequency band a few seconds to minutes prior to their merger. 
Therefore, we have selected three typical SBBHs with equal mass ratios and total masses of $3\ \mathrm{M_\odot }$, $20\ \mathrm{M_\odot }$, and $100\ \mathrm{M_\odot }$, respectively.
The waveforms for the 10 minutes preceding ISCO are computed.
For MBHB sources observed by space-based detectors, we select MBHBs with three representative masses: $10^5\ \mathrm{M_\odot }$, $10^6\ \mathrm{M_\odot }$, and $10^7\ \mathrm{M_\odot }$. 
The waveforms are calculated for the three months before ISCO. 
For multiband observations, the SBBHs with masses of $20\ \mathrm{M_\odot }$ and $100\ \mathrm{M_\odot }$ are extended to the low-frequency band, lasting for a year before reaching 0.1 Hz.
The SBBHs beyond this threshold would take approximately three months ($20\ \mathrm{M_\odot }$) and six days ($100\ \mathrm{M_\odot }$) to enter the frequency band of ground-based detectors, providing an opportunity for conducting multiband observations.

\begin{table}[ht]
\centering
\renewcommand{\arraystretch}{1.5}
\caption{Parameter distribution used in calculation~\cite{my_paper3}. $U[a,b]$ represents a uniform distribution from $a$ to $b$.}\label{tab:parameters}
\begin{tabular*}{\columnwidth}{@{\extracolsep{\fill}}lr@{}}
\hline
 Parameter & Distribution \\
\hline
Ecliptic longitude $\lambda$ & $U[0,2\pi]$ rad\\
Ecliptic latitude $\beta$ & $\arcsin(U[-1,1])$ rad\\
Equatorial longitude $\alpha$ & $U[0,2\pi]$ rad\\
Equatorial latitude $\delta$ & $\arcsin(U[-1,1])$ rad\\
Polarization $\psi$ & $U[0,2\pi]$ rad\\
Initial phase $\phi_0$ & $U[0,2\pi]$ rad\\
\hline
\end{tabular*}
\end{table}

We generate 100 BBH sources. 
The sky positions for BBH sources are selected with identical parameters, following the distribution in Table~\ref{tab:parameters}.
In order to investigate the impact of inclination angle, we consider 45 angles, resulting in a total of 4500 calculations of SNR and FIM for each BBH source.
The corresponding results are presented in Secs.~\ref{sec:Results_inclination} and \ref{sec:Results_ppE}. 
Finally, we set the fiducial value for the ppE parameter to $10^{-4}$ (cf.~\cite{ppE_Taiji}).

\section{Constraints on parameters}\label{sec:Constraints_parameters}
\subsection{Results with inclination}\label{sec:Results_inclination}
\begin{figure}[ht]
    \begin{minipage}{\columnwidth}
        \centering
        \includegraphics[width=0.8\textwidth,
        trim=0 0 0 0,clip]{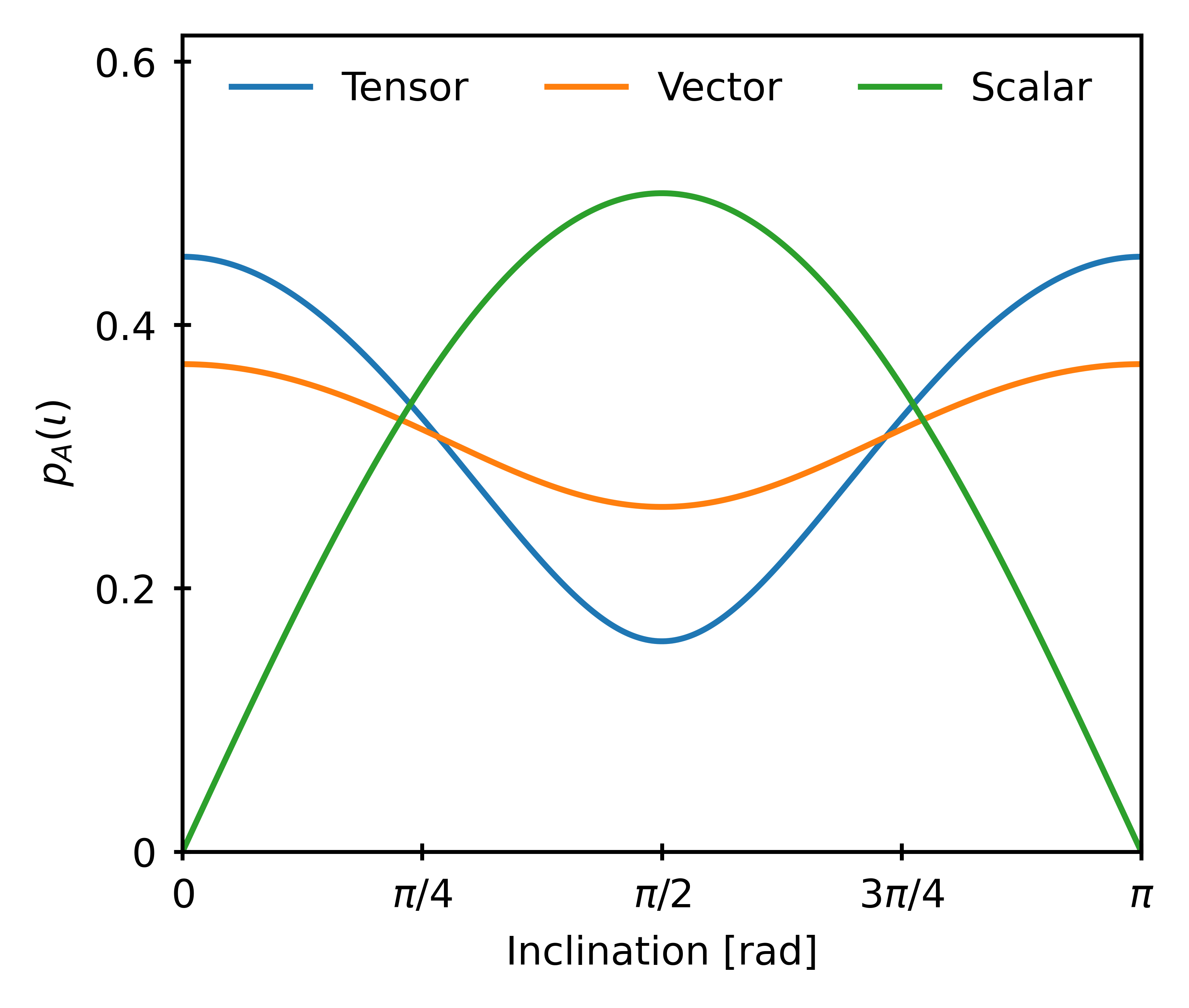}
        \caption{The distributions of the inclination for tensor mode $p_T(\iota)$, vector mode $p_V(\iota)$ and scalar mode $p_S(\iota)$. All distributions have been normalized.}\label{fig:pA(l)}
    \end{minipage}
\end{figure}

The relationship between ppE parameter uncertainty and the inclination angle is specifically examined. 
By referring to Eq.~(\ref{eq:waveform}), we can define the distribution of inclination for the tensor mode as
\begin{equation}
    p_T(\iota)\propto \sqrt{\left((1+\cos^2\iota)/2\right)^2+\left(\cos\iota \right)^2} .
\end{equation}
The distributions of vector mode $p_V(\iota)$ and scalar mode $p_S(\iota)$ can be determined similarly,
\begin{equation}
    \begin{aligned}
p_V(\iota)&\propto \sqrt{\left(\cos\iota \right)^2+1^2}, \\
p_S(\iota)&\propto \sqrt{\left(\sin\iota \right)^2+\left(\sin\iota \right)^2}.
\end{aligned}
\end{equation}
The distributions of these three modes are presented in Fig.~\ref{fig:pA(l)}.

\begin{figure}[ht]
    \begin{minipage}{\columnwidth}
        \centering
        \includegraphics[width=0.95\textwidth,
        trim=0 0 0 0,clip]{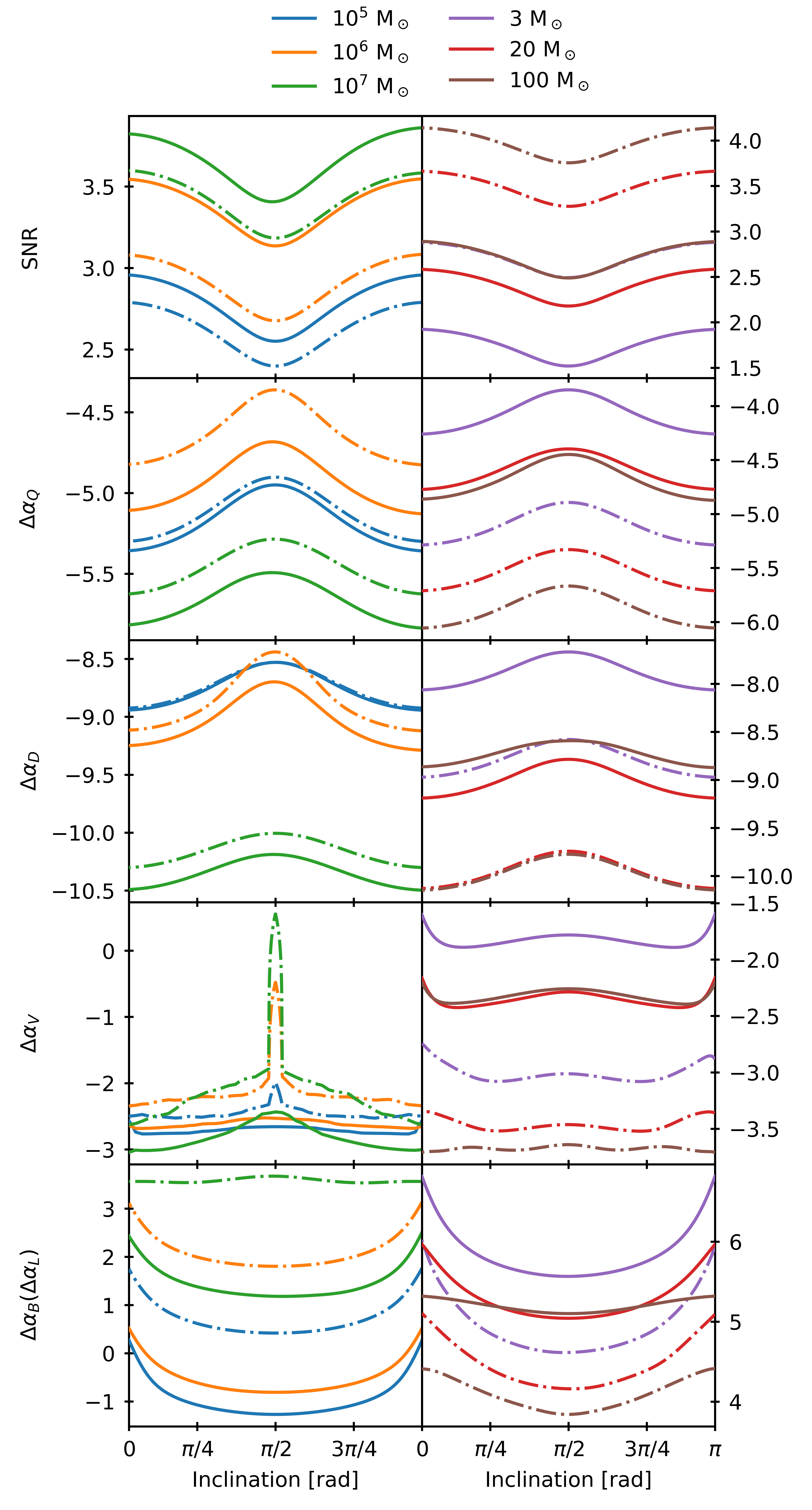}
        \caption{Relationship between SNR and parameter uncertainty with inclination angle variation. The vertical axis is represented in logarithmic scale. The SNR is calculated based on the GR case, and each line in the graph represents the median value. The results for $\Delta \alpha_B$ and $\Delta \alpha_L$ are identical, therefore only one is shown. The left panel includes LISA (solid line) and TianQin (dashdot line), while the right panel includes LIGO (solid line) and ET (dashdot line).}\label{fig:result_inclination}
    \end{minipage}
\end{figure}

The distribution of the scalar mode with inclination angle, as shown in Fig.~\ref{fig:pA(l)}, exhibits an opposite trend to that of tensor and vector modes. 
It reaches its maximum value at $\iota=\pi/2$ and minimum values at $\iota=0,\pi$. 
The scalar mode outperforms the other two modes at $\iota=\pi/2$, while approaching zero at $\iota=0,\pi$. 
The patterns of variation in both tensor and vector modes are similar, although tensor modes show a greater amplitude of change. 
The SNR and FIM have been computed to determine the uncertainty of ppE parameters, as shown in Fig.~\ref{fig:result_inclination}.

The scalar mode in Fig.~\ref{fig:result_inclination} shows an opposite trend compared with other modes, as depicted in the distribution shown in Fig.~\ref{fig:pA(l)}. 
The scalar mode vanishes at $\iota=0,\pi$, resulting in a near absence of signal around these values. 
As a result, the parameter uncertainty for the scalar mode approaches infinity in Fig.~\ref{fig:result_inclination}.
Additionally, considering the low-frequency limit, a $10^5\ \mathrm{M_\odot } $ MBHB with low-SNR demonstrates smaller $\Delta \alpha_B$ and $\Delta \alpha_L$ values compared with a $10^7\ \mathrm{M_\odot } $ MBHB with high-SNR. 
Owing to superior sensitivity and the lower transfer frequency, LISA exhibits significantly better $\Delta \alpha_B$ and $\Delta \alpha_L$ values than TianQin does. 
Ground-based detectors are unable to distinguish between the breathing mode and longitudinal mode; hence their corresponding $\Delta \alpha_B$ and $\Delta \alpha_L$ values are much larger than those of space-based detectors and become invalid.

Besides the scalar mode, a higher SNR indicates a more robust signal with reduced parameter uncertainties. 
In our calculations, SNR is computed based on the GR case. 
Therefore, the SNR value is primarily influenced by tensor modes. 
In contrast to tensor and scalar modes, the uncertainty of the vector mode is significantly higher at $\iota=\pi/2$ for TianQin. 
The fixed orientation of TianQin limits variations in the response function at that specific angle, resulting in a significant increase in $\Delta \alpha_V$ near $\iota=\pi/2$.

The transfer frequency not only directly impacts the degeneracy between breathing and longitudinal modes but also affects other aspects.
The transfer function $\mathcal{T}$ is directly related to frequency, providing constraints beyond the low-frequency limit.
$\alpha_Q$ and $\alpha_D$ determine the variation of the orbital angular frequency, and the influence of $\mathcal{T}$ can reduce parameter uncertainty.
Therefore, the $\Delta \alpha_Q$ of $10^5\ \mathrm{M_\odot } $ MBHB is smaller than that for a $10^6\ \mathrm{M_\odot } $ MBHB, and $\Delta \alpha_D$ for both MBHBs also varies with the relationship near $\iota=\pi/2$. 
Because of the low-frequency limit, ground-based detectors are not affected by $\mathcal{T}$, causing the uncertainties $\Delta \alpha_Q$ and $\Delta \alpha_D$ for SBBHs that depend solely on SNR.
The aim of this section is to analyze the correlation between ppE parameter uncertainty and inclination angle for two detectors, one space-based and one ground-based.

\subsection{Results with ppE parameters}\label{sec:Results_ppE}
\begin{figure*}[ht]
    \begin{minipage}{\textwidth}
        \centering
        \includegraphics[width=0.9\textwidth,
        trim=0 0 0 0,clip]{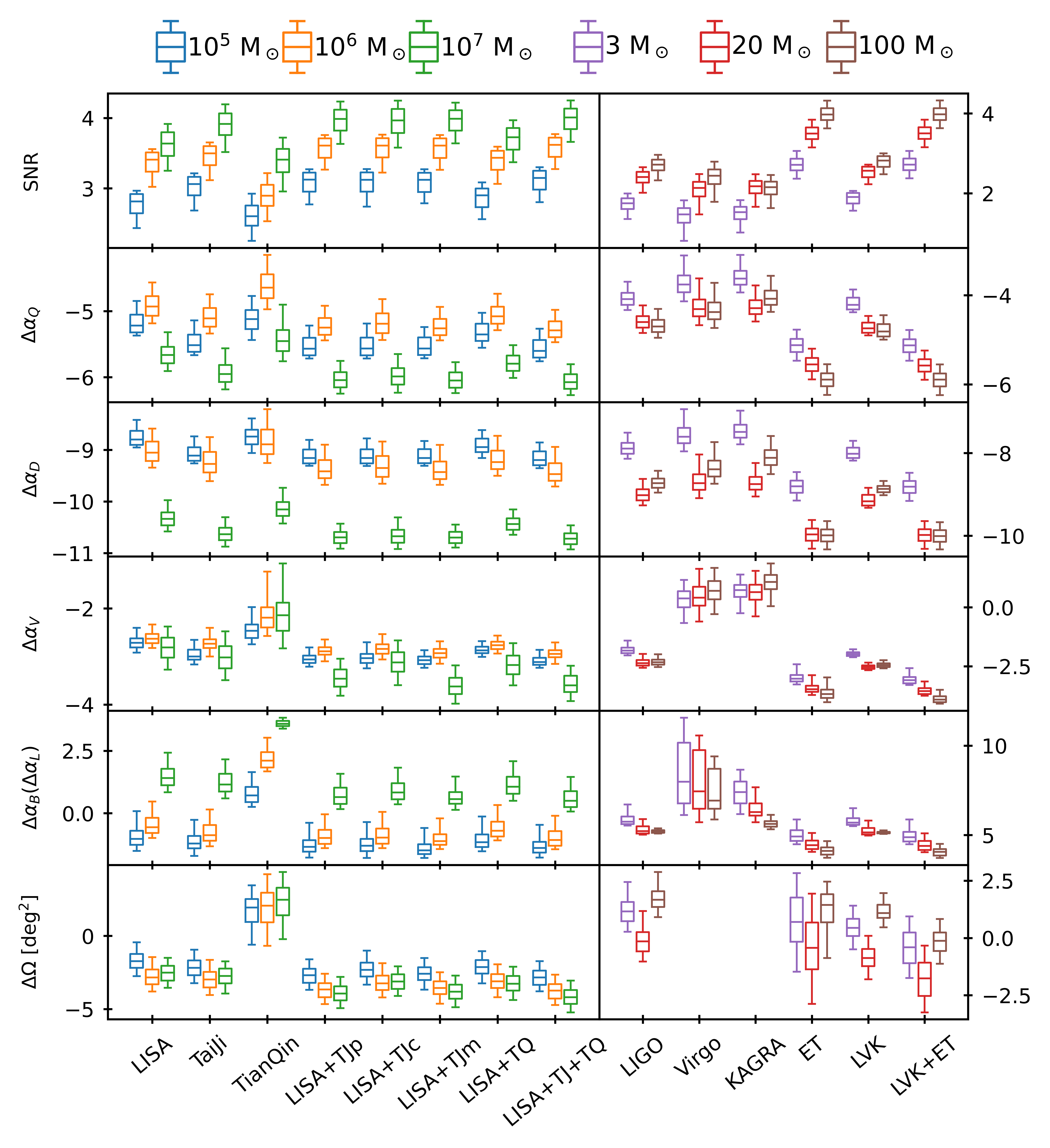}
        \caption{SNR and parameter uncertainty with space- and ground-based detectors. The vertical axis is represented in logarithmic scale. The upper and lower horizontal lines in each box represent the 90\% confidence interval. The edges of the box correspond to the upper and lower quartiles, while the lines inside the box represent the median. The uncertainty of sky position is expressed as $\Delta \Omega = 2\pi|\sin\theta_e|(\Sigma_{\theta_e\theta_e}\Sigma_{\phi_e\phi_e} - \Sigma^2_{\theta_e\phi_e})^{1/2}$~\cite{Omega}. The left panel shows the results for the space-based detectors, where TJ represents Taiji and TQ represents TianQin. In our calculations, there is no difference between the two alternative orbital configurations for TianQin, so they are not labeled separately. The right panel displays the results for the ground-based detectors, but due to the extremely poor positioning capabilities of individual Virgo and KAGRA detectors, their $\Delta \Omega$ values are not shown.}\label{fig:result_ppE}
    \end{minipage}
\end{figure*}

The performance of several typical BBHs is assessed using different detectors and various parameters.
The results for different detectors and their networks are depicted in Fig.~\ref{fig:result_ppE}.

Regarding space-based detectors, Taiji, which shares a similar configuration with LISA, surpasses LISA across all aspects, demonstrating superior SNR and reduced parameter uncertainty.
TianQin's unique orbital configuration results in inferior performance compared with LISA and Taiji.  
Additionally, the significant increase in TianQin's parameter uncertainty at $\iota=\pi/2$, particularly in $\Delta \alpha_V$, causes the upper limit of the box plot for $\Delta \alpha_V$ to be substantially higher than its median value as shown in the comprehensive results presented in Fig.~\ref{fig:result_ppE}.

When considering the space-based detector network, significant improvements are observed across all metrics compared with individual detectors.
Due to the heliocentric orbits for both LISA and Taiji, their network configuration remains stable, ensuring a consistent angle between the detectors.
The three alternative orbital configurations of Taiji, along with LISA, exhibit distinct angles that introduce outcome variations. 
A larger angular separation between the detectors increases sky area coverage with high sensitivity, resulting in differences in parameter uncertainty under similar SNR conditions.
The LISA+TJm configuration demonstrates superior performance overall, with LISA+TJp outperforming LISA+TJc, particularly in parameters such as $\Delta \alpha_V$ and $\Delta \Omega$. 
Furthermore, it is evident that LISA+TJ+TQ surpasses LISA+TJ, highlighting the advantage of a network comprising three detectors over two. 
For a more comprehensive analysis of space detector networks, please refer to Ref.~\cite{my_paper2}.

The ability of ground-based detectors like Virgo and KAGRA to detect additional polarization and determine the sky position is limited.
This limitation stems from the fact that a single detector only has only one response angle, and a 10-minute GW signal is very short compared with the observational period (Earth's rotation period).
In the ground-based detector's coordinate system, a 10-minute SBBH source travels only $2.5^\circ$, whereas a 90-day MBHB source moves nearly $90^\circ$ in the space-based detector coordinate system.
Consequently, the parameter uncertainty range for ground-based detectors is larger.
Furthermore, both LIGO and ET have three detectors, but ET's detectors are all coplanar while LIGO's are not. 
The relative positions of LIGO's three detectors offer greater advantages compared with ET.
Despite having an SNR one order of magnitude higher than LIGO, ET still has very similar $\Delta \Omega$ values.

Regarding the ground-based detector network, the currently operational LVK is unable to compensate for the sensitivity differences between second-generation and third-generation detectors. 
The limitations of ppE parameters in LVK differ by one to two orders of magnitude compared with ET. 
Furthermore, due to LVK's varied response angles, $\Delta \Omega$ values are slightly superior to those of ET.
The combination of LVK+ET provides the most comprehensive enhancement in network sensitivity.
For instance, $\Delta \alpha_Q$ and $\Delta \alpha_D$ can approach the sensitivity level achieved by space-based detector networks, with potential surpassing in $\Delta \alpha_V$. 
Due to its arm length, the ground-based detector network's $\Delta \alpha_B$ and $\Delta \alpha_L$ are four to six orders of magnitude higher than those observed in space-based detector networks, making it impossible to distinguish between breathing and longitudinal modes.

\subsection{Results with luminosity distance}\label{sec:Results_distance}
The distances we assumed in our previous calculations and analysis were fixed: $D_L=44.6$ Mpc for SBBH and $D_L=6.79$ Gpc for MBHB.
The GW signal can vary due to different distances, potentially affecting the final results.
From Eq.~(\ref{eq:waveform}), the strain is inversely proportional to the luminosity distance, represented as $h\propto 1/D_L$.
Regarding the ppE parameter $\xi$, the partial derivative of the strain with respect to it, $\partial h/\partial\xi \propto 1/D_L$. 
Consequently, the FIM is $(\partial h/\partial\xi|\partial h/\partial\xi )\propto 1/D_L^2$, and thus the uncertainty of the ppE parameter scales as $\Delta \xi \propto D_L$.
In other words, the uncertainty of the ppE parameter is directly proportional to the distance, 
\begin{equation}\label{eq:k}
    \Delta \xi(D_L) =\mathcal{K}D_L+\Delta \xi(0),
\end{equation}
where $\mathcal{K}$ is a proportionality constant. 
The results for SBBH and MBHB are calculated at various distances, as presented in Table~\ref{tab:k}.
\begin{table}[ht]
\centering
\renewcommand{\arraystretch}{1.5}
\caption{Proportional coefficient $\mathcal{K}$ corresponding to different ppE parameters. We use logarithmic $\log(\mathcal{K} )$ to represent the results in the table.}\label{tab:k}
\begin{tabular*}{\columnwidth}{@{\extracolsep{\fill}}cccccc@{}}
\hline
Detector & $M[\mathrm{M_\odot } ]$ & $\Delta\alpha_Q$ & $\Delta\alpha_D$ & $\Delta\alpha_V$ & $\Delta\alpha_B(\alpha_L)$\\
\hline
 & $10^{5}$ & -9.05 & -12.63 & -6.55 & -4.86 \\ 
LISA &  $10^{6} $ & -8.76 & -12.89 & -6.46 & -4.39 \\ 
 & $10^{7} $ & -9.50 & -14.17 & -6.64 & -2.42 \\ 
\hline
 & $10^{5}$ & -9.35 & -12.94 & -6.83 & -5.05 \\ 
TaiJi &  $10^{6} $ & -8.94 & -13.11 & -6.57 & -4.71 \\ 
 & $10^{7} $ & -9.78 & -14.46 & -6.85 & -2.68 \\ 
\hline
 & $10^{5}$ & -8.95 & -12.58 & -6.30 & -3.12 \\ 
TianQin &  $10^{6} $ & -8.47 & -12.73 & -6.03 & -1.72 \\ 
 & $10^{7} $ & -9.28 & -13.99 & -5.97 & -0.25 \\ 
\hline
 & $3$ & -5.74 & -9.54 & -3.49 & 4.10 \\ 
LIGO &  $20$ & -6.25 & -10.67 & -4.02 & 3.57 \\ 
 & $100$ & -6.35 & -10.38 & -3.98 & 3.54 \\ 
\hline
 & $3$ & -5.41 & -9.26 & -1.27 & 6.25 \\ 
Virgo &  $20$ & -5.96 & -10.38 & -1.24 & 5.75 \\ 
 & $100$ & -6.03 & -10.04 & -0.96 & 5.35 \\ 
\hline
 & $3$ & -5.28 & -9.14 & -0.92 & 5.76 \\ 
KAGRA &  $20$ & -5.94 & -10.40 & -1.01 & 4.61 \\ 
 & $100$ & -5.73 & -9.77 & -0.58 & 3.97 \\ 
\hline
 & $3$ & -6.77 & -10.45 & -4.68 & 3.24 \\ 
ET &  $20$ & -7.20 & -11.62 & -5.13 & 2.79 \\ 
 & $100$ & -7.54 & -11.64 & -5.32 & 2.46 \\ 
\hline
\end{tabular*}
\end{table}

Equation~(\ref{eq:k}) reveals that the uncertainty of the ppE parameters is directly proportional to the distance.
The rate of change in uncertainty with distance varies for different ppE parameters. 
Table~\ref{tab:k} illustrates that the influence of distance on $\Delta \alpha_Q$ and $\Delta \alpha_D$ is negligible, since most $\mathcal{K}$ values are less than $10^{-5}$.
This can be attributed to the fact that these two parameters are closely linked to frequency, and any alteration in GW amplitude caused by distance has an negligible effect on frequency resolution. 
Moreover, ground-based detectors exhibit relatively larger $\mathcal{K}$ values compared with space-based detectors.

The $\mathcal{K}$ for $\Delta \alpha_V$ exhibits a significantly larger magnitude than the first two parameters.
The $\mathcal{K}$ values of space-based detectors are typically around $10^{-6}$, while Virgo and KAGRA have $\mathcal{K}$ values around $10^{-1}$.
LIGO and ET outperform Virgo and KAGRA, in most cases with $\mathcal{K} < 10^{-3}$. 
These observations are consistent the analysis presented in the previous section, which elucidates these differences stem from detector performance.

Due to the degeneracy in ground-based detectors, $\Delta \alpha_B$ ($\Delta \alpha_L$) exhibits significantly large values of $\mathcal{K}$, exceeding $10^2$.
In contrast, space-based detectors effectively resolve this degeneracy and improves outcomes.
The results in Table~\ref{tab:k} depend on the frequency range of MBHBs and the transfer frequency.
The frequencies of smaller mass MBHBs are lower and closer to the transfer frequency, enhancing the resolution of these two modes and resulting in smaller $\mathcal{K}$ values.
The results are consistent across different detectors: Taiji exhibits the lowest transfer frequency, whereas TianQin demonstrates the highest, aligning with their respective $\mathcal{K}$ values.

\subsection{Multiband observation}\label{sec:Multiband_observation}
\begin{figure}[ht]
    \begin{minipage}{\columnwidth}
        \centering
        \includegraphics[width=0.95\textwidth,
        trim=0 0 0 0,clip]{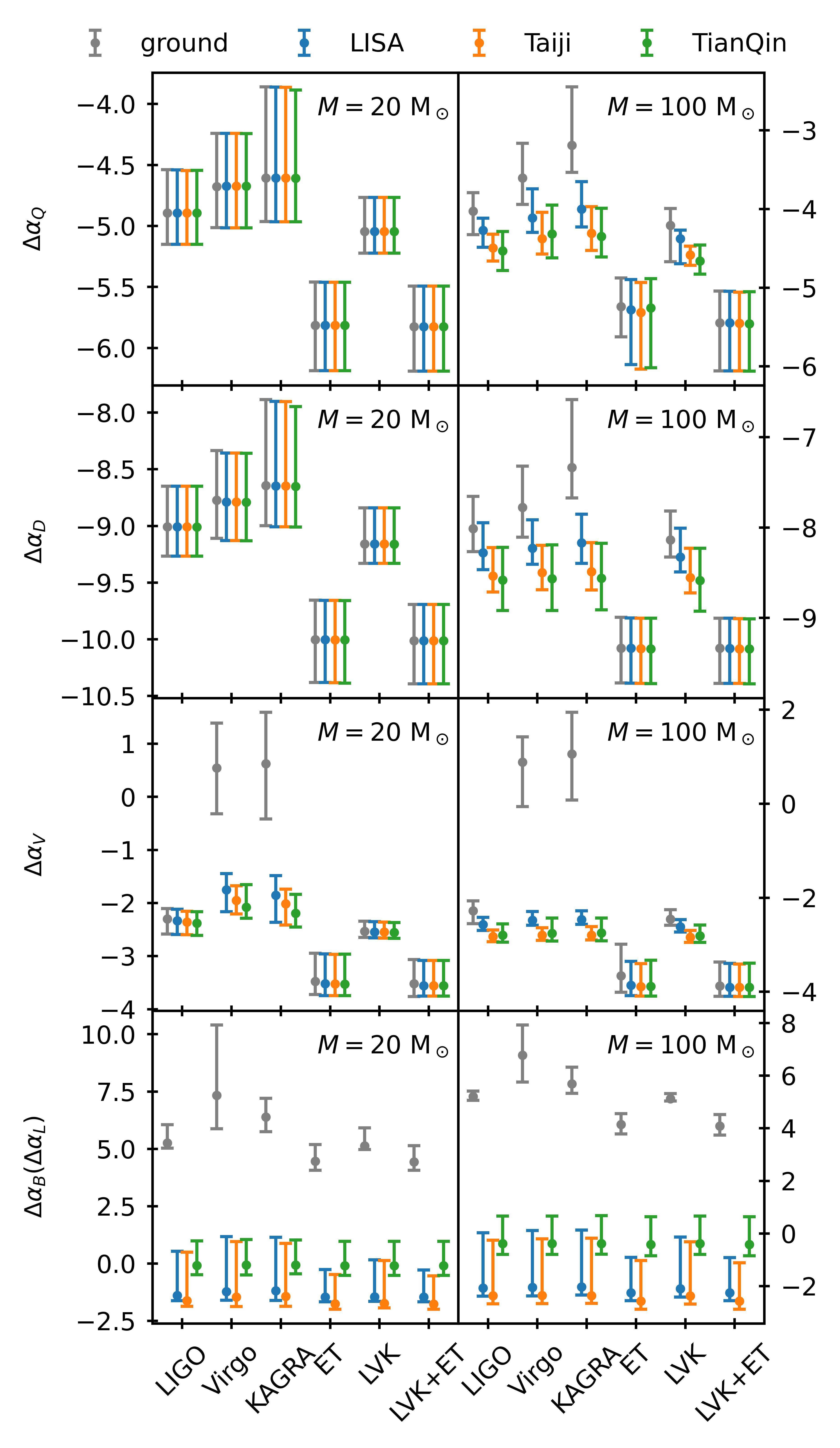}
        \caption{The uncertainty of ppE parameters in multiband observations. The vertical axis is represented in logarithmic scale. For simplicity, we use points to represent the median, with error bars above and below indicating the 90\% confidence interval. In the left panel, the total mass of the SBBH source is $M=20\ \mathrm{M_\odot }$, while in the right panel, it is $M=100\ \mathrm{M_\odot }$. To save computational resources, only a subset of the sources are calculated, resulting in slight differences from Fig.~\ref{fig:result_ppE}.}\label{fig:multiband}
    \end{minipage}
\end{figure}

Space-based detectors can observe both MBHBs and SBBHs.
Among numerous sources, only a few SBBHs rapidly increase in frequency and merge into the frequency band of ground-based detectors, enabling multiband observations by both space- and ground-based detectors~\cite{LISA_SBBH}.
The FIM is calculated for two types of SBBHs observed across multiple frequency bands and compared with results solely obtained from ground-based detectors, as shown in Fig.~\ref{fig:multiband}.

Multiband observations greatly enhances the capabilities of ground-based detectors, particularly in effectively distinguishing between breathing and longitudinal modes.
The addition of space-based detectors resolves the degeneracy, enabling multiband observations of SBBH to constrain $\Delta \alpha_B$ and $\Delta \alpha_L$ at levels comparable with those achieved by observing MBHB with space-based detectors.
Additionally, the multiband observations greatly enhance the performance of Virgo and KAGRA in $\Delta \alpha_V$, reducing it by nearly three orders of magnitude. 
This brings them closer to the levels achieved by LIGO and ET.
The $\Delta \alpha_V$ level of the ground-based detector network remains higher than that of the space-based detector network, even with multiband observations.
Therefore, while there is improvement from these observations, its significance may not be as pronounced.

The improvements for SBBH due to multiband observations vary with the mass, with respect to $\Delta \alpha_Q$ and $\Delta \alpha_D$.
For SBBH with $M=20\ \mathrm{M_\odot }$, there is no significant improvement observed, while for a mass of $100\ \mathrm{M_\odot }$, some enhancement is demonstrated.
The sensitivity of ET is so high that multiband observation does not significantly improve ET and LVK+ET, but it notably enhances the performance of only second-generation detectors.
The frequency ranges of both types of SBBH differ within the space-based detector band, despite their one-year duration.
The frequency variation is 46 mHz for SBBHs with $M=20\ \mathrm{M_\odot}$ and 78 mHz for those with $M=100\ \mathrm{M_\odot }$.
The stronger constraints are caused by greater frequency variations, which in turn affect the observed improvements. 
Additionally, larger frequency variations occur with longer observation times.
While four-year observations generally outperform one-year observations, this improvement is not significant. 
This is because the further away from the merger, the smaller the frequency change over the same time period. 
The SBBH with $M=100\ \mathrm{M_\odot}$ experiences a frequency shift of 78 mHz within one year before reaching 0.1 Hz.
The change over four years was only 87 mHz, with the additional change within three years being just 9 mHz.
Hence, our choice of a one-year observation duration remains justifiable.

The performance of various space-based detectors significantly impacts the outcomes of multi-band observations.
Moreover, the cutoff frequency of SBBHs within the space-based detector band is chosen to be 0.1 Hz, which is lower than TianQin's transfer frequency of 0.28 Hz. 
That disparity renders TianQin less effective than LISA and Taiji in observing breathing and longitudinal modes, resulting in a much weaker multiband observation improvement for $\Delta \alpha_B$ and $\Delta \alpha_L$.
Conversely, TianQin exhibits significantly enhanced multiband observations for $\Delta \alpha_Q$, $\Delta \alpha_D$, and $\Delta \alpha_V$ due to its superior sensitivity within the selected SBBH frequency band compared with other detectors.
The inclusion of space-based detectors compensates for the degeneracy observed in ground-based detectors regarding scalar modes and enhances other aspects.
The advancement allows SBBH observation results to reach a level comparable to MBHB results.

\subsection{Multimessenger observation}
Apart from using space- and ground-based detectors for multiband observations, integrating multimessenger observations, including electromagnetic (EM) observations, can also enhance GW detections.
EM observations provide a distinctive viewpoint on BBH sources that is separate from GW observations. 
The studies conducted by Refs.~\cite{EM_MBHB1,EM_MBHB2} have demonstrated that mergers of MBHBs can emit EM radiation from accretion disks. 
Additionally, the mergers of SBBHs in active galactic nuclei have the potential to generate powerful EM jets~\cite{EM_MBHB1,EM_MBHB2,EM_SBBH1,EM_SBBH2}.
These findings highlight the significance of both scenarios as promising targets for EM detectors.
Accurate determination of the source's sky position facilitates subsequent EM follow-up observations to identify counterparts. 
The findings from EM observations can then serve as valuable prior information to minimize uncertainties in GW observations.

The EM effects generated by BBH mergers can be detected by infrared, optical, and x-ray observatories, with diverse implications for advancing GW research.
For simplicity, we consider an ideal scenario where EM observations accurately determine specific parameters, disregarding discrepancies among EM detectors. 
According to Refs.~\cite{TianQinGBs,my_paper2}, the FIM removes the corresponding row and column when incorporating EM observations in order to mitigate uncertainties of other parameters in the GW data. 
This approach evaluates the performance enhancement of GW observations under ideal multimessenger conditions.
The enhancements are quantified when EM observations accurately determine the distance, sky position, or inclination angle, and these results are compared with GW observations alone. 
Notably, although this study only focuse on ppE parameters, multimessenger observations also enhance time, mass, polarization, and other parameters.  
For details on enhancements in non-ppE parameters, please refer to Ref.~\cite{my_paper2}.

The multimessenger observations for LISA and Taiji demonstrate significant enhancements primarily for MBHBs with $M=10^7\ \mathrm{M_\odot}$.
Utilizing multimessenger observations to determine the distance or inclination angle can result in a reduction of $\Delta \alpha_V$, $\Delta \alpha_B$, and $\Delta \alpha_L$ by 15\% to 24\%.
The determination of the sky position can lead to a reduction in these parameters ranging from 36\% to 49\%.
The multimessenger observations significantly enhance TianQin's detection capabilities for all three typical MBHB masses.
When accurately determining the distance, inclination angle, or sky position, $\Delta \alpha_V$ can be reduced by 40\% to 57\%. 
Furthermore, reductions in $\Delta \alpha_B$ and $\Delta \alpha_L$ are less, ranging from 9\% to 17\%.

The impact of multimessenger observations on $\Delta \alpha_B$ and $\Delta \alpha_L$ is significantly diminished for ground-based detectors compared with multiband observations, as it fails to resolve the degeneracy and thus is not further considered.
The improvement shown by LIGO is relatively less significant than that of ET, mainly because of its varied response angles. 
The determination of the distance or inclination angle leads to a reduction in $\Delta \alpha_V$ by 5\% to 7\%, while negligible changes are observed when determining the sky position.
The enhancements exhibited by ET are more significant.
When any one of the three parameters is determined, $\Delta \alpha_V$ can be reduced by 15\% to 43\%, with greater reductions observed for higher-mass SBBHs.
Moreover, multimessenger observations significantly enhance Virgo and KAGRA, reducing $\Delta \alpha_V$ by 97\% to 99.4\%. 
The improvement achieved is only one order of magnitude lower than that observed in LIGO, demonstrating a significant advancement beyond GW observations alone.

Multimessenger observations, whether for space- or ground-based detectors, do not improve the other two ppE parameters, $\Delta \alpha_Q$ and $\Delta \alpha_D$. 
The reason is that these two ppE parameters directly link to the highest frequency of GWs, where precision in observations is at its peak.
Theoretically, multimessenger observations could enhance constraints on all ppE parameters, but in practice, some improvements only offer marginal benefits.
The results presented in this study are limited to cases where the parameter uncertainty is reduced by more than 5\%, with most unreported results reducing less than 1\%. 
In conclusion, multimessenger observations can greatly improve the effectiveness of GW observations, resulting in stronger constraints on ppE parameters.

\section{Conclusion}\label{sec:Conclusion}
In this paper, we investigate the expected targets for constraining GW polarization using space- and ground-based detectors within the ppE framework. 
Specifically, we adopt a model-independent ppE framework that incorporates a GW waveform with all six polarization modes, adhering to current GR test results. 
The space-based detectors considered in this study include LISA, Taiji, and TianQin, along with alternative orbital configurations for individual use or in network scenarios.
As for the ground-based detectors, we examine the currently operational LVK and the third-generation detector ET, evaluating them both individually and in network configurations.
Our analysis focuses on the performance of detectors across different polarization modes, with an emphasis on the perspective of the response function. 
Additionally, we employ the FIM to quantify ppE parameter constraints for both space- and ground-based detectors by simulating three typical masses of MBHB and SBBH. 
We present the results under various combinations and further explore how multiband and multimessenger observations enhance ppE constraints, providing a comprehensive analysis from diverse perspectives.

\textbf{For space-based detectors,} Taiji provides the most stringent constraints on GW polarization modes, with LISA outperforming TianQin.  
Specifically, TianQin exhibits significantly weaker constraints on scalar modes compared with LISA and Taiji, but the uncertainties in other ppE parameters are similar between the two detectors.
In network scenarios, the combination of LISA+TJm slightly surpasses other combinations in performance, and a network comprising all three detectors outperforms networks consisting of only two detectors.

\textbf{For ground-based detectors,} the differences in constraints on frequency-related ppE parameters are negligible, while LIGO's results for vector modes outperform those of Virgo and KAGRA, approaching the performance level of ET. 
In network scenarios, the LVK combination does not surpass the results of ET alone, with the LVK+ET network providing the most optimal constraints.. 
Overall, ground-based detectors lack discrimination between breathing and longitudinal modes and exhibit significantly lower capability in constraining scalar modes compared with space detectors. 
Apart from this distinction, the constraints on ppE parameters from both space- and ground-based detectors are comparable.

\textbf{For multiband observation,} it can effectively resolves the degeneracies encountered by ground-based detectors in breathing and longitudinal modes, addressing their inability to discern scalar modes. 
This approach also enhances the capability of Virgo and KAGRA to constrain vector modes, reaching a level comparable to LIGO and offering certain improvements for other ppE parameters. 
Furthermore, TianQin exhibits the most significant enhancement, followed by Taiji, with LISA demonstrating comparatively lesser improvement.

\textbf{For multimessenger observation,} it provides the most significant enhancement for TianQin, bringing its results for vector modes close to those of LISA. 
Regarding scalar modes, both LISA and Taiji witness greater improvements. 
ET demonstrates a larger improvement compared with LIGO, while Virgo and KAGRA experience substantial enhancements with their results for vector modes approaching the level of LIGO. 
Overall, both multiband and multimessenger observations can augment GW detection outcomes to a certain extent, leading to improved constraints on GW polarization modes.

The future research will be approached from multiple perspectives.
Firstly, we will utilize longer-duration GW signals, which possess a greater capacity for information content. 
Additionally, we will consider the propagation of GWs at speeds other than that of light due to different modes. 
It is worth noting that vector and scalar modes may arrive before or simultaneously with tensor modes~\cite{GW_speed}.
Secondly, our objective is to delve into the effects brought about by time-delay interferometry (TDI) technology. 
This entails exploring the subtle differences in outcomes resulting from various TDI combinations~\cite{TDI}. 
Moreover, we will examine diverse sources and assess the impact of ppE parameters on BBH systems with unequal masses.
Thirdly, our plan involves employing specific modified theories of gravity in conjunction with existing GW events to yield more stringent constraint results~\cite{non_GR_event1,non_GR_event2,non_GR_event3,Einstein_Cartan1,Einstein_Cartan2,Einstein_Cartan3}.
Lastly, within the context of multimessenger observations, we aim to investigate the performance of EM detectors under realistic conditions. 
This approach allows us to provide results that align more closely with actual scenarios and enhance GW observations.
The aim of these in-depth studies is to advance the research on GW polarization modes, comprehensively and meticulously evaluate detector performance, and provide more valuable results.

\begin{acknowledgements}
This work was supported by the National Key Research and Development Program of China (Grant No. 2023YFC2206702), the National Natural Science Foundation of China (Grant No. 12347101), and the Natural Science Foundation of Chongqing (Grant No. CSTB2023NSCQ-MSX0103). 
\end{acknowledgements}

\bibliography{references}

%apsrev4-2.bst 2019-01-14 (MD) hand-edited version of apsrev4-1.bst
%Control: key (0)
%Control: author (8) initials jnrlst
%Control: editor formatted (1) identically to author
%Control: production of article title (0) allowed
%Control: page (0) single
%Control: year (1) truncated
%Control: production of eprint (0) enabled
\begin{thebibliography}{84}%
\makeatletter
\providecommand \@ifxundefined [1]{%
 \@ifx{#1\undefined}
}%
\providecommand \@ifnum [1]{%
 \ifnum #1\expandafter \@firstoftwo
 \else \expandafter \@secondoftwo
 \fi
}%
\providecommand \@ifx [1]{%
 \ifx #1\expandafter \@firstoftwo
 \else \expandafter \@secondoftwo
 \fi
}%
\providecommand \natexlab [1]{#1}%
\providecommand \enquote  [1]{``#1''}%
\providecommand \bibnamefont  [1]{#1}%
\providecommand \bibfnamefont [1]{#1}%
\providecommand \citenamefont [1]{#1}%
\providecommand \href@noop [0]{\@secondoftwo}%
\providecommand \href [0]{\begingroup \@sanitize@url \@href}%
\providecommand \@href[1]{\@@startlink{#1}\@@href}%
\providecommand \@@href[1]{\endgroup#1\@@endlink}%
\providecommand \@sanitize@url [0]{\catcode `\\12\catcode `\$12\catcode `\&12\catcode `\#12\catcode `\^12\catcode `\_12\catcode `\%12\relax}%
\providecommand \@@startlink[1]{}%
\providecommand \@@endlink[0]{}%
\providecommand \url  [0]{\begingroup\@sanitize@url \@url }%
\providecommand \@url [1]{\endgroup\@href {#1}{\urlprefix }}%
\providecommand \urlprefix  [0]{URL }%
\providecommand \Eprint [0]{\href }%
\providecommand \doibase [0]{https://doi.org/}%
\providecommand \selectlanguage [0]{\@gobble}%
\providecommand \bibinfo  [0]{\@secondoftwo}%
\providecommand \bibfield  [0]{\@secondoftwo}%
\providecommand \translation [1]{[#1]}%
\providecommand \BibitemOpen [0]{}%
\providecommand \bibitemStop [0]{}%
\providecommand \bibitemNoStop [0]{.\EOS\space}%
\providecommand \EOS [0]{\spacefactor3000\relax}%
\providecommand \BibitemShut  [1]{\csname bibitem#1\endcsname}%
\let\auto@bib@innerbib\@empty
%</preamble>
\bibitem [{\citenamefont {Stairs}(2003)}]{test_GR1}%
  \BibitemOpen
  \bibfield  {author} {\bibinfo {author} {\bibfnamefont {I.~H.}\ \bibnamefont {Stairs}},\ }\bibfield  {title} {\bibinfo {title} {{Testing general relativity with pulsar timing}},\ }\href {https://doi.org/10.12942/lrr-2003-5} {\bibfield  {journal} {\bibinfo  {journal} {Living Rev. Rel.}\ }\textbf {\bibinfo {volume} {6}},\ \bibinfo {pages} {5} (\bibinfo {year} {2003})},\ \Eprint {https://arxiv.org/abs/astro-ph/0307536} {arXiv:astro-ph/0307536} \BibitemShut {NoStop}%
\bibitem [{\citenamefont {Will}(2014)}]{test_GR2}%
  \BibitemOpen
  \bibfield  {author} {\bibinfo {author} {\bibfnamefont {C.~M.}\ \bibnamefont {Will}},\ }\bibfield  {title} {\bibinfo {title} {{The Confrontation between General Relativity and Experiment}},\ }\href {https://doi.org/10.12942/lrr-2014-4} {\bibfield  {journal} {\bibinfo  {journal} {Living Rev. Rel.}\ }\textbf {\bibinfo {volume} {17}},\ \bibinfo {pages} {4} (\bibinfo {year} {2014})},\ \Eprint {https://arxiv.org/abs/1403.7377} {arXiv:1403.7377 [gr-qc]} \BibitemShut {NoStop}%
\bibitem [{\citenamefont {Baker}\ \emph {et~al.}(2015)\citenamefont {Baker}, \citenamefont {Psaltis},\ and\ \citenamefont {Skordis}}]{test_GR3}%
  \BibitemOpen
  \bibfield  {author} {\bibinfo {author} {\bibfnamefont {T.}~\bibnamefont {Baker}}, \bibinfo {author} {\bibfnamefont {D.}~\bibnamefont {Psaltis}},\ and\ \bibinfo {author} {\bibfnamefont {C.}~\bibnamefont {Skordis}},\ }\bibfield  {title} {\bibinfo {title} {{Linking Tests of Gravity On All Scales: from the Strong-Field Regime to Cosmology}},\ }\href {https://doi.org/10.1088/0004-637X/802/1/63} {\bibfield  {journal} {\bibinfo  {journal} {Astrophys. J.}\ }\textbf {\bibinfo {volume} {802}},\ \bibinfo {pages} {63} (\bibinfo {year} {2015})},\ \Eprint {https://arxiv.org/abs/1412.3455} {arXiv:1412.3455 [astro-ph.CO]} \BibitemShut {NoStop}%
\bibitem [{\citenamefont {Clifton}\ \emph {et~al.}(2012)\citenamefont {Clifton}, \citenamefont {Ferreira}, \citenamefont {Padilla},\ and\ \citenamefont {Skordis}}]{test_GR4}%
  \BibitemOpen
  \bibfield  {author} {\bibinfo {author} {\bibfnamefont {T.}~\bibnamefont {Clifton}}, \bibinfo {author} {\bibfnamefont {P.~G.}\ \bibnamefont {Ferreira}}, \bibinfo {author} {\bibfnamefont {A.}~\bibnamefont {Padilla}},\ and\ \bibinfo {author} {\bibfnamefont {C.}~\bibnamefont {Skordis}},\ }\bibfield  {title} {\bibinfo {title} {{Modified Gravity and Cosmology}},\ }\href {https://doi.org/10.1016/j.physrep.2012.01.001} {\bibfield  {journal} {\bibinfo  {journal} {Phys. Rept.}\ }\textbf {\bibinfo {volume} {513}},\ \bibinfo {pages} {1} (\bibinfo {year} {2012})},\ \Eprint {https://arxiv.org/abs/1106.2476} {arXiv:1106.2476 [astro-ph.CO]} \BibitemShut {NoStop}%
\bibitem [{\citenamefont {Berti}\ \emph {et~al.}(2015)\citenamefont {Berti} \emph {et~al.}}]{test_GR5}%
  \BibitemOpen
  \bibfield  {author} {\bibinfo {author} {\bibfnamefont {E.}~\bibnamefont {Berti}} \emph {et~al.},\ }\bibfield  {title} {\bibinfo {title} {{Testing General Relativity with Present and Future Astrophysical Observations}},\ }\href {https://doi.org/10.1088/0264-9381/32/24/243001} {\bibfield  {journal} {\bibinfo  {journal} {Class. Quant. Grav.}\ }\textbf {\bibinfo {volume} {32}},\ \bibinfo {pages} {243001} (\bibinfo {year} {2015})},\ \Eprint {https://arxiv.org/abs/1501.07274} {arXiv:1501.07274 [gr-qc]} \BibitemShut {NoStop}%
\bibitem [{\citenamefont {Shi}\ \emph {et~al.}(2023)\citenamefont {Shi}, \citenamefont {Ji}, \citenamefont {Zhang},\ and\ \citenamefont {Mei}}]{ppE_TianQin}%
  \BibitemOpen
  \bibfield  {author} {\bibinfo {author} {\bibfnamefont {C.}~\bibnamefont {Shi}}, \bibinfo {author} {\bibfnamefont {M.}~\bibnamefont {Ji}}, \bibinfo {author} {\bibfnamefont {J.-d.}\ \bibnamefont {Zhang}},\ and\ \bibinfo {author} {\bibfnamefont {J.}~\bibnamefont {Mei}},\ }\bibfield  {title} {\bibinfo {title} {{Testing general relativity with TianQin: The prospect of using the inspiral signals of black hole binaries}},\ }\href {https://doi.org/10.1103/PhysRevD.108.024030} {\bibfield  {journal} {\bibinfo  {journal} {Phys. Rev. D}\ }\textbf {\bibinfo {volume} {108}},\ \bibinfo {pages} {024030} (\bibinfo {year} {2023})},\ \Eprint {https://arxiv.org/abs/2210.13006} {arXiv:2210.13006 [gr-qc]} \BibitemShut {NoStop}%
\bibitem [{\citenamefont {Eardley}\ \emph {et~al.}(1973)\citenamefont {Eardley}, \citenamefont {Lee}, \citenamefont {Lightman}, \citenamefont {Wagoner},\ and\ \citenamefont {Will}}]{GR_polarization}%
  \BibitemOpen
  \bibfield  {author} {\bibinfo {author} {\bibfnamefont {D.~M.}\ \bibnamefont {Eardley}}, \bibinfo {author} {\bibfnamefont {D.~L.}\ \bibnamefont {Lee}}, \bibinfo {author} {\bibfnamefont {A.~P.}\ \bibnamefont {Lightman}}, \bibinfo {author} {\bibfnamefont {R.~V.}\ \bibnamefont {Wagoner}},\ and\ \bibinfo {author} {\bibfnamefont {C.~M.}\ \bibnamefont {Will}},\ }\bibfield  {title} {\bibinfo {title} {{Gravitational-wave observations as a tool for testing relativistic gravity}},\ }\href {https://doi.org/10.1103/PhysRevLett.30.884} {\bibfield  {journal} {\bibinfo  {journal} {Phys. Rev. Lett.}\ }\textbf {\bibinfo {volume} {30}},\ \bibinfo {pages} {884} (\bibinfo {year} {1973})}\BibitemShut {NoStop}%
\bibitem [{\citenamefont {Brans}\ and\ \citenamefont {Dicke}(1961)}]{BD_polarization}%
  \BibitemOpen
  \bibfield  {author} {\bibinfo {author} {\bibfnamefont {C.}~\bibnamefont {Brans}}\ and\ \bibinfo {author} {\bibfnamefont {R.~H.}\ \bibnamefont {Dicke}},\ }\bibfield  {title} {\bibinfo {title} {{Mach's principle and a relativistic theory of gravitation}},\ }\href {https://doi.org/10.1103/PhysRev.124.925} {\bibfield  {journal} {\bibinfo  {journal} {Phys. Rev.}\ }\textbf {\bibinfo {volume} {124}},\ \bibinfo {pages} {925} (\bibinfo {year} {1961})}\BibitemShut {NoStop}%
\bibitem [{\citenamefont {Alves}\ \emph {et~al.}(2009)\citenamefont {Alves}, \citenamefont {Miranda},\ and\ \citenamefont {de~Araujo}}]{fR_polarization1}%
  \BibitemOpen
  \bibfield  {author} {\bibinfo {author} {\bibfnamefont {M.~E.~S.}\ \bibnamefont {Alves}}, \bibinfo {author} {\bibfnamefont {O.~D.}\ \bibnamefont {Miranda}},\ and\ \bibinfo {author} {\bibfnamefont {J.~C.~N.}\ \bibnamefont {de~Araujo}},\ }\bibfield  {title} {\bibinfo {title} {{Probing the f(R) formalism through gravitational wave polarizations}},\ }\href {https://doi.org/10.1016/j.physletb.2009.08.005} {\bibfield  {journal} {\bibinfo  {journal} {Phys. Lett. B}\ }\textbf {\bibinfo {volume} {679}},\ \bibinfo {pages} {401} (\bibinfo {year} {2009})},\ \Eprint {https://arxiv.org/abs/0908.0861} {arXiv:0908.0861 [gr-qc]} \BibitemShut {NoStop}%
\bibitem [{\citenamefont {Rizwana~Kausar}\ \emph {et~al.}(2016)\citenamefont {Rizwana~Kausar}, \citenamefont {Philippoz},\ and\ \citenamefont {Jetzer}}]{fR_polarization2}%
  \BibitemOpen
  \bibfield  {author} {\bibinfo {author} {\bibfnamefont {H.}~\bibnamefont {Rizwana~Kausar}}, \bibinfo {author} {\bibfnamefont {L.}~\bibnamefont {Philippoz}},\ and\ \bibinfo {author} {\bibfnamefont {P.}~\bibnamefont {Jetzer}},\ }\bibfield  {title} {\bibinfo {title} {{Gravitational Wave Polarization Modes in $f(R)$ Theories}},\ }\href {https://doi.org/10.1103/PhysRevD.93.124071} {\bibfield  {journal} {\bibinfo  {journal} {Phys. Rev. D}\ }\textbf {\bibinfo {volume} {93}},\ \bibinfo {pages} {124071} (\bibinfo {year} {2016})},\ \Eprint {https://arxiv.org/abs/1606.07000} {arXiv:1606.07000 [gr-qc]} \BibitemShut {NoStop}%
\bibitem [{\citenamefont {Kobayashi}(2019)}]{Horndeski_polarization}%
  \BibitemOpen
  \bibfield  {author} {\bibinfo {author} {\bibfnamefont {T.}~\bibnamefont {Kobayashi}},\ }\bibfield  {title} {\bibinfo {title} {{Horndeski theory and beyond: a review}},\ }\href {https://doi.org/10.1088/1361-6633/ab2429} {\bibfield  {journal} {\bibinfo  {journal} {Rept. Prog. Phys.}\ }\textbf {\bibinfo {volume} {82}},\ \bibinfo {pages} {086901} (\bibinfo {year} {2019})},\ \Eprint {https://arxiv.org/abs/1901.07183} {arXiv:1901.07183 [gr-qc]} \BibitemShut {NoStop}%
\bibitem [{\citenamefont {Jacobson}\ and\ \citenamefont {Mattingly}(2004)}]{Einstein-Aether_polarization1}%
  \BibitemOpen
  \bibfield  {author} {\bibinfo {author} {\bibfnamefont {T.}~\bibnamefont {Jacobson}}\ and\ \bibinfo {author} {\bibfnamefont {D.}~\bibnamefont {Mattingly}},\ }\bibfield  {title} {\bibinfo {title} {{Einstein-Aether waves}},\ }\href {https://doi.org/10.1103/PhysRevD.70.024003} {\bibfield  {journal} {\bibinfo  {journal} {Phys. Rev. D}\ }\textbf {\bibinfo {volume} {70}},\ \bibinfo {pages} {024003} (\bibinfo {year} {2004})},\ \Eprint {https://arxiv.org/abs/gr-qc/0402005} {arXiv:gr-qc/0402005} \BibitemShut {NoStop}%
\bibitem [{\citenamefont {Lin}\ \emph {et~al.}(2019)\citenamefont {Lin}, \citenamefont {Zhao}, \citenamefont {Zhang}, \citenamefont {Liu}, \citenamefont {Wang}, \citenamefont {Zhang}, \citenamefont {Zhang}, \citenamefont {Zhao}, \citenamefont {Zhu},\ and\ \citenamefont {Wang}}]{Einstein-Aether_polarization2}%
  \BibitemOpen
  \bibfield  {author} {\bibinfo {author} {\bibfnamefont {K.}~\bibnamefont {Lin}}, \bibinfo {author} {\bibfnamefont {X.}~\bibnamefont {Zhao}}, \bibinfo {author} {\bibfnamefont {C.}~\bibnamefont {Zhang}}, \bibinfo {author} {\bibfnamefont {T.}~\bibnamefont {Liu}}, \bibinfo {author} {\bibfnamefont {B.}~\bibnamefont {Wang}}, \bibinfo {author} {\bibfnamefont {S.}~\bibnamefont {Zhang}}, \bibinfo {author} {\bibfnamefont {X.}~\bibnamefont {Zhang}}, \bibinfo {author} {\bibfnamefont {W.}~\bibnamefont {Zhao}}, \bibinfo {author} {\bibfnamefont {T.}~\bibnamefont {Zhu}},\ and\ \bibinfo {author} {\bibfnamefont {A.}~\bibnamefont {Wang}},\ }\bibfield  {title} {\bibinfo {title} {{Gravitational waveforms, polarizations, response functions, and energy losses of triple systems in Einstein-aether theory}},\ }\href {https://doi.org/10.1103/PhysRevD.99.023010} {\bibfield  {journal} {\bibinfo  {journal} {Phys. Rev. D}\ }\textbf {\bibinfo {volume} {99}},\ \bibinfo {pages} {023010} (\bibinfo {year} {2019})},\ \Eprint
  {https://arxiv.org/abs/1810.07707} {arXiv:1810.07707 [astro-ph.GA]} \BibitemShut {NoStop}%
\bibitem [{\citenamefont {Bekenstein}(2004)}]{TVS_polarization1}%
  \BibitemOpen
  \bibfield  {author} {\bibinfo {author} {\bibfnamefont {J.~D.}\ \bibnamefont {Bekenstein}},\ }\bibfield  {title} {\bibinfo {title} {{Relativistic gravitation theory for the MOND paradigm}},\ }\href {https://doi.org/10.1103/PhysRevD.70.083509} {\bibfield  {journal} {\bibinfo  {journal} {Phys. Rev. D}\ }\textbf {\bibinfo {volume} {70}},\ \bibinfo {pages} {083509} (\bibinfo {year} {2004})},\ \bibinfo {note} {[Erratum: Phys.Rev.D 71, 069901 (2005)]},\ \Eprint {https://arxiv.org/abs/astro-ph/0403694} {arXiv:astro-ph/0403694} \BibitemShut {NoStop}%
\bibitem [{\citenamefont {Gong}\ and\ \citenamefont {Hou}(2018)}]{TVS_polarization2}%
  \BibitemOpen
  \bibfield  {author} {\bibinfo {author} {\bibfnamefont {Y.}~\bibnamefont {Gong}}\ and\ \bibinfo {author} {\bibfnamefont {S.}~\bibnamefont {Hou}},\ }\bibfield  {title} {\bibinfo {title} {{The Polarizations of Gravitational Waves}},\ }\href {https://doi.org/10.3390/universe4080085} {\bibfield  {journal} {\bibinfo  {journal} {Universe}\ }\textbf {\bibinfo {volume} {4}},\ \bibinfo {pages} {85} (\bibinfo {year} {2018})},\ \Eprint {https://arxiv.org/abs/1806.04027} {arXiv:1806.04027 [gr-qc]} \BibitemShut {NoStop}%
\bibitem [{\citenamefont {Abbott}\ \emph {et~al.}(2016{\natexlab{a}})\citenamefont {Abbott} \emph {et~al.}}]{GW150914}%
  \BibitemOpen
  \bibfield  {author} {\bibinfo {author} {\bibfnamefont {B.~P.}\ \bibnamefont {Abbott}} \emph {et~al.} (\bibinfo {collaboration} {LIGO Scientific, Virgo}),\ }\bibfield  {title} {\bibinfo {title} {{Observation of Gravitational Waves from a Binary Black Hole Merger}},\ }\href {https://doi.org/10.1103/PhysRevLett.116.061102} {\bibfield  {journal} {\bibinfo  {journal} {Phys. Rev. Lett.}\ }\textbf {\bibinfo {volume} {116}},\ \bibinfo {pages} {061102} (\bibinfo {year} {2016}{\natexlab{a}})},\ \Eprint {https://arxiv.org/abs/1602.03837} {arXiv:1602.03837 [gr-qc]} \BibitemShut {NoStop}%
\bibitem [{\citenamefont {Abbott}\ \emph {et~al.}(2019{\natexlab{a}})\citenamefont {Abbott} \emph {et~al.}}]{GWTC1}%
  \BibitemOpen
  \bibfield  {author} {\bibinfo {author} {\bibfnamefont {B.~P.}\ \bibnamefont {Abbott}} \emph {et~al.} (\bibinfo {collaboration} {LIGO Scientific, Virgo}),\ }\bibfield  {title} {\bibinfo {title} {{GWTC-1: A Gravitational-Wave Transient Catalog of Compact Binary Mergers Observed by LIGO and Virgo during the First and Second Observing Runs}},\ }\href {https://doi.org/10.1103/PhysRevX.9.031040} {\bibfield  {journal} {\bibinfo  {journal} {Phys. Rev. X}\ }\textbf {\bibinfo {volume} {9}},\ \bibinfo {pages} {031040} (\bibinfo {year} {2019}{\natexlab{a}})},\ \Eprint {https://arxiv.org/abs/1811.12907} {arXiv:1811.12907 [astro-ph.HE]} \BibitemShut {NoStop}%
\bibitem [{\citenamefont {Abbott}\ \emph {et~al.}(2021{\natexlab{a}})\citenamefont {Abbott} \emph {et~al.}}]{GWTC2}%
  \BibitemOpen
  \bibfield  {author} {\bibinfo {author} {\bibfnamefont {R.}~\bibnamefont {Abbott}} \emph {et~al.} (\bibinfo {collaboration} {LIGO Scientific, Virgo}),\ }\bibfield  {title} {\bibinfo {title} {{GWTC-2: Compact Binary Coalescences Observed by LIGO and Virgo During the First Half of the Third Observing Run}},\ }\href {https://doi.org/10.1103/PhysRevX.11.021053} {\bibfield  {journal} {\bibinfo  {journal} {Phys. Rev. X}\ }\textbf {\bibinfo {volume} {11}},\ \bibinfo {pages} {021053} (\bibinfo {year} {2021}{\natexlab{a}})},\ \Eprint {https://arxiv.org/abs/2010.14527} {arXiv:2010.14527 [gr-qc]} \BibitemShut {NoStop}%
\bibitem [{\citenamefont {Abbott}\ \emph {et~al.}(2023)\citenamefont {Abbott} \emph {et~al.}}]{GWTC3}%
  \BibitemOpen
  \bibfield  {author} {\bibinfo {author} {\bibfnamefont {R.}~\bibnamefont {Abbott}} \emph {et~al.} (\bibinfo {collaboration} {LIGO Scientific, Virgo}),\ }\href@noop {} {\bibinfo {title} {{GWTC-3: Compact Binary Coalescences Observed by LIGO and Virgo During the Second Part of the Third Observing Run}}} (\bibinfo {year} {2023}),\ \Eprint {https://arxiv.org/abs/2111.03606} {arXiv:2111.03606 [gr-qc]} \BibitemShut {NoStop}%
\bibitem [{\citenamefont {Abbott}\ \emph {et~al.}(2019{\natexlab{b}})\citenamefont {Abbott} \emph {et~al.}}]{GW_test_GR1}%
  \BibitemOpen
  \bibfield  {author} {\bibinfo {author} {\bibfnamefont {B.~P.}\ \bibnamefont {Abbott}} \emph {et~al.} (\bibinfo {collaboration} {LIGO Scientific, Virgo}),\ }\bibfield  {title} {\bibinfo {title} {{Tests of General Relativity with the Binary Black Hole Signals from the LIGO-Virgo Catalog GWTC-1}},\ }\href {https://doi.org/10.1103/PhysRevD.100.104036} {\bibfield  {journal} {\bibinfo  {journal} {Phys. Rev. D}\ }\textbf {\bibinfo {volume} {100}},\ \bibinfo {pages} {104036} (\bibinfo {year} {2019}{\natexlab{b}})},\ \Eprint {https://arxiv.org/abs/1903.04467} {arXiv:1903.04467 [gr-qc]} \BibitemShut {NoStop}%
\bibitem [{\citenamefont {Abbott}\ \emph {et~al.}(2021{\natexlab{b}})\citenamefont {Abbott} \emph {et~al.}}]{GW_test_GR2}%
  \BibitemOpen
  \bibfield  {author} {\bibinfo {author} {\bibfnamefont {R.}~\bibnamefont {Abbott}} \emph {et~al.} (\bibinfo {collaboration} {LIGO Scientific, Virgo}),\ }\bibfield  {title} {\bibinfo {title} {{Tests of general relativity with binary black holes from the second LIGO-Virgo gravitational-wave transient catalog}},\ }\href {https://doi.org/10.1103/PhysRevD.103.122002} {\bibfield  {journal} {\bibinfo  {journal} {Phys. Rev. D}\ }\textbf {\bibinfo {volume} {103}},\ \bibinfo {pages} {122002} (\bibinfo {year} {2021}{\natexlab{b}})},\ \Eprint {https://arxiv.org/abs/2010.14529} {arXiv:2010.14529 [gr-qc]} \BibitemShut {NoStop}%
\bibitem [{\citenamefont {Collaboration}\ \emph {et~al.}(2021)\citenamefont {Collaboration}, \citenamefont {the Virgo~Collaboration},\ and\ \citenamefont {the KAGRA~Collaboration}}]{GW_test_GR3}%
  \BibitemOpen
  \bibfield  {author} {\bibinfo {author} {\bibfnamefont {T.~L.~S.}\ \bibnamefont {Collaboration}}, \bibinfo {author} {\bibnamefont {the Virgo~Collaboration}},\ and\ \bibinfo {author} {\bibnamefont {the KAGRA~Collaboration}},\ }\href@noop {} {\bibinfo {title} {Tests of general relativity with gwtc-3}} (\bibinfo {year} {2021}),\ \Eprint {https://arxiv.org/abs/2112.06861} {arXiv:2112.06861 [gr-qc]} \BibitemShut {NoStop}%
\bibitem [{\citenamefont {Abbott}\ \emph {et~al.}(2016{\natexlab{b}})\citenamefont {Abbott} \emph {et~al.}}]{GW_test_GR4}%
  \BibitemOpen
  \bibfield  {author} {\bibinfo {author} {\bibfnamefont {B.~P.}\ \bibnamefont {Abbott}} \emph {et~al.} (\bibinfo {collaboration} {LIGO Scientific, Virgo}),\ }\bibfield  {title} {\bibinfo {title} {{Tests of general relativity with GW150914}},\ }\href {https://doi.org/10.1103/PhysRevLett.116.221101} {\bibfield  {journal} {\bibinfo  {journal} {Phys. Rev. Lett.}\ }\textbf {\bibinfo {volume} {116}},\ \bibinfo {pages} {221101} (\bibinfo {year} {2016}{\natexlab{b}})},\ \bibinfo {note} {[Erratum: Phys.Rev.Lett. 121, 129902 (2018)]},\ \Eprint {https://arxiv.org/abs/1602.03841} {arXiv:1602.03841 [gr-qc]} \BibitemShut {NoStop}%
\bibitem [{\citenamefont {Abbott}\ \emph {et~al.}(2019{\natexlab{c}})\citenamefont {Abbott} \emph {et~al.}}]{GW_test_GR5}%
  \BibitemOpen
  \bibfield  {author} {\bibinfo {author} {\bibfnamefont {B.~P.}\ \bibnamefont {Abbott}} \emph {et~al.} (\bibinfo {collaboration} {LIGO Scientific, Virgo}),\ }\bibfield  {title} {\bibinfo {title} {{Tests of General Relativity with GW170817}},\ }\href {https://doi.org/10.1103/PhysRevLett.123.011102} {\bibfield  {journal} {\bibinfo  {journal} {Phys. Rev. Lett.}\ }\textbf {\bibinfo {volume} {123}},\ \bibinfo {pages} {011102} (\bibinfo {year} {2019}{\natexlab{c}})},\ \Eprint {https://arxiv.org/abs/1811.00364} {arXiv:1811.00364 [gr-qc]} \BibitemShut {NoStop}%
\bibitem [{\citenamefont {Berti}\ \emph {et~al.}(2018{\natexlab{a}})\citenamefont {Berti}, \citenamefont {Yagi},\ and\ \citenamefont {Yunes}}]{GW_test_GR6}%
  \BibitemOpen
  \bibfield  {author} {\bibinfo {author} {\bibfnamefont {E.}~\bibnamefont {Berti}}, \bibinfo {author} {\bibfnamefont {K.}~\bibnamefont {Yagi}},\ and\ \bibinfo {author} {\bibfnamefont {N.}~\bibnamefont {Yunes}},\ }\bibfield  {title} {\bibinfo {title} {{Extreme Gravity Tests with Gravitational Waves from Compact Binary Coalescences: (I) Inspiral-Merger}},\ }\href {https://doi.org/10.1007/s10714-018-2362-8} {\bibfield  {journal} {\bibinfo  {journal} {Gen. Rel. Grav.}\ }\textbf {\bibinfo {volume} {50}},\ \bibinfo {pages} {46} (\bibinfo {year} {2018}{\natexlab{a}})},\ \Eprint {https://arxiv.org/abs/1801.03208} {arXiv:1801.03208 [gr-qc]} \BibitemShut {NoStop}%
\bibitem [{\citenamefont {Berti}\ \emph {et~al.}(2018{\natexlab{b}})\citenamefont {Berti}, \citenamefont {Yagi}, \citenamefont {Yang},\ and\ \citenamefont {Yunes}}]{GW_test_GR7}%
  \BibitemOpen
  \bibfield  {author} {\bibinfo {author} {\bibfnamefont {E.}~\bibnamefont {Berti}}, \bibinfo {author} {\bibfnamefont {K.}~\bibnamefont {Yagi}}, \bibinfo {author} {\bibfnamefont {H.}~\bibnamefont {Yang}},\ and\ \bibinfo {author} {\bibfnamefont {N.}~\bibnamefont {Yunes}},\ }\bibfield  {title} {\bibinfo {title} {{Extreme Gravity Tests with Gravitational Waves from Compact Binary Coalescences: (II) Ringdown}},\ }\href {https://doi.org/10.1007/s10714-018-2372-6} {\bibfield  {journal} {\bibinfo  {journal} {Gen. Rel. Grav.}\ }\textbf {\bibinfo {volume} {50}},\ \bibinfo {pages} {49} (\bibinfo {year} {2018}{\natexlab{b}})},\ \Eprint {https://arxiv.org/abs/1801.03587} {arXiv:1801.03587 [gr-qc]} \BibitemShut {NoStop}%
\bibitem [{\citenamefont {Liang}\ \emph {et~al.}(2019)\citenamefont {Liang}, \citenamefont {Gong}, \citenamefont {Weinstein}, \citenamefont {Zhang},\ and\ \citenamefont {Zhang}}]{frequency_response}%
  \BibitemOpen
  \bibfield  {author} {\bibinfo {author} {\bibfnamefont {D.}~\bibnamefont {Liang}}, \bibinfo {author} {\bibfnamefont {Y.}~\bibnamefont {Gong}}, \bibinfo {author} {\bibfnamefont {A.~J.}\ \bibnamefont {Weinstein}}, \bibinfo {author} {\bibfnamefont {C.}~\bibnamefont {Zhang}},\ and\ \bibinfo {author} {\bibfnamefont {C.}~\bibnamefont {Zhang}},\ }\bibfield  {title} {\bibinfo {title} {{Frequency response of space-based interferometric gravitational-wave detectors}},\ }\href {https://doi.org/10.1103/PhysRevD.99.104027} {\bibfield  {journal} {\bibinfo  {journal} {Phys. Rev. D}\ }\textbf {\bibinfo {volume} {99}},\ \bibinfo {pages} {104027} (\bibinfo {year} {2019})},\ \Eprint {https://arxiv.org/abs/1901.09624} {arXiv:1901.09624 [gr-qc]} \BibitemShut {NoStop}%
\bibitem [{\citenamefont {Aasi}\ \emph {et~al.}(2015)\citenamefont {Aasi} \emph {et~al.}}]{aLIGO}%
  \BibitemOpen
  \bibfield  {author} {\bibinfo {author} {\bibfnamefont {J.}~\bibnamefont {Aasi}} \emph {et~al.} (\bibinfo {collaboration} {LIGO Scientific}),\ }\bibfield  {title} {\bibinfo {title} {{Advanced LIGO}},\ }\href {https://doi.org/10.1088/0264-9381/32/7/074001} {\bibfield  {journal} {\bibinfo  {journal} {Class. Quant. Grav.}\ }\textbf {\bibinfo {volume} {32}},\ \bibinfo {pages} {074001} (\bibinfo {year} {2015})},\ \Eprint {https://arxiv.org/abs/1411.4547} {arXiv:1411.4547 [gr-qc]} \BibitemShut {NoStop}%
\bibitem [{\citenamefont {Acernese}\ \emph {et~al.}(2015)\citenamefont {Acernese} \emph {et~al.}}]{aVIRGO}%
  \BibitemOpen
  \bibfield  {author} {\bibinfo {author} {\bibfnamefont {F.}~\bibnamefont {Acernese}} \emph {et~al.} (\bibinfo {collaboration} {VIRGO}),\ }\bibfield  {title} {\bibinfo {title} {{Advanced Virgo: a second-generation interferometric gravitational wave detector}},\ }\href {https://doi.org/10.1088/0264-9381/32/2/024001} {\bibfield  {journal} {\bibinfo  {journal} {Class. Quant. Grav.}\ }\textbf {\bibinfo {volume} {32}},\ \bibinfo {pages} {024001} (\bibinfo {year} {2015})},\ \Eprint {https://arxiv.org/abs/1408.3978} {arXiv:1408.3978 [gr-qc]} \BibitemShut {NoStop}%
\bibitem [{\citenamefont {Somiya}(2012)}]{KAGRA}%
  \BibitemOpen
  \bibfield  {author} {\bibinfo {author} {\bibfnamefont {K.}~\bibnamefont {Somiya}} (\bibinfo {collaboration} {KAGRA}),\ }\bibfield  {title} {\bibinfo {title} {{Detector configuration of KAGRA: The Japanese cryogenic gravitational-wave detector}},\ }\href {https://doi.org/10.1088/0264-9381/29/12/124007} {\bibfield  {journal} {\bibinfo  {journal} {Class. Quant. Grav.}\ }\textbf {\bibinfo {volume} {29}},\ \bibinfo {pages} {124007} (\bibinfo {year} {2012})},\ \Eprint {https://arxiv.org/abs/1111.7185} {arXiv:1111.7185 [gr-qc]} \BibitemShut {NoStop}%
\bibitem [{\citenamefont {Amaro-Seoane}\ \emph {et~al.}(2017)\citenamefont {Amaro-Seoane}, \citenamefont {Audley}, \citenamefont {Babak}, \citenamefont {Baker}, \citenamefont {Barausse}, \citenamefont {Bender}, \citenamefont {Berti}, \citenamefont {Binetruy}, \citenamefont {Born}, \citenamefont {Bortoluzzi} \emph {et~al.}}]{LISA}%
  \BibitemOpen
  \bibfield  {author} {\bibinfo {author} {\bibfnamefont {P.}~\bibnamefont {Amaro-Seoane}}, \bibinfo {author} {\bibfnamefont {H.}~\bibnamefont {Audley}}, \bibinfo {author} {\bibfnamefont {S.}~\bibnamefont {Babak}}, \bibinfo {author} {\bibfnamefont {J.}~\bibnamefont {Baker}}, \bibinfo {author} {\bibfnamefont {E.}~\bibnamefont {Barausse}}, \bibinfo {author} {\bibfnamefont {P.}~\bibnamefont {Bender}}, \bibinfo {author} {\bibfnamefont {E.}~\bibnamefont {Berti}}, \bibinfo {author} {\bibfnamefont {P.}~\bibnamefont {Binetruy}}, \bibinfo {author} {\bibfnamefont {M.}~\bibnamefont {Born}}, \bibinfo {author} {\bibfnamefont {D.}~\bibnamefont {Bortoluzzi}}, \emph {et~al.},\ }\bibfield  {title} {\bibinfo {title} {Laser interferometer space antenna},\ }\href@noop {} {\bibfield  {journal} {\bibinfo  {journal} {arXiv preprint arXiv:1702.00786}\ } (\bibinfo {year} {2017})}\BibitemShut {NoStop}%
\bibitem [{\citenamefont {Hu}\ and\ \citenamefont {Wu}(2017)}]{Taiji}%
  \BibitemOpen
  \bibfield  {author} {\bibinfo {author} {\bibfnamefont {W.-R.}\ \bibnamefont {Hu}}\ and\ \bibinfo {author} {\bibfnamefont {Y.-L.}\ \bibnamefont {Wu}},\ }\bibfield  {title} {\bibinfo {title} {{The Taiji Program in Space for gravitational wave physics and the nature of gravity}},\ }\href {https://doi.org/10.1093/nsr/nwx116} {\bibfield  {journal} {\bibinfo  {journal} {Natl. Sci. Rev.}\ }\textbf {\bibinfo {volume} {4}},\ \bibinfo {pages} {685} (\bibinfo {year} {2017})}\BibitemShut {NoStop}%
\bibitem [{\citenamefont {Luo}\ \emph {et~al.}(2016)\citenamefont {Luo} \emph {et~al.}}]{TianQin}%
  \BibitemOpen
  \bibfield  {author} {\bibinfo {author} {\bibfnamefont {J.}~\bibnamefont {Luo}} \emph {et~al.} (\bibinfo {collaboration} {TianQin}),\ }\bibfield  {title} {\bibinfo {title} {{TianQin: a space-borne gravitational wave detector}},\ }\href {https://doi.org/10.1088/0264-9381/33/3/035010} {\bibfield  {journal} {\bibinfo  {journal} {Class. Quant. Grav.}\ }\textbf {\bibinfo {volume} {33}},\ \bibinfo {pages} {035010} (\bibinfo {year} {2016})},\ \Eprint {https://arxiv.org/abs/1512.02076} {arXiv:1512.02076 [astro-ph.IM]} \BibitemShut {NoStop}%
\bibitem [{\citenamefont {Gair}\ \emph {et~al.}(2013)\citenamefont {Gair}, \citenamefont {Vallisneri}, \citenamefont {Larson},\ and\ \citenamefont {Baker}}]{test_GR_space}%
  \BibitemOpen
  \bibfield  {author} {\bibinfo {author} {\bibfnamefont {J.~R.}\ \bibnamefont {Gair}}, \bibinfo {author} {\bibfnamefont {M.}~\bibnamefont {Vallisneri}}, \bibinfo {author} {\bibfnamefont {S.~L.}\ \bibnamefont {Larson}},\ and\ \bibinfo {author} {\bibfnamefont {J.~G.}\ \bibnamefont {Baker}},\ }\bibfield  {title} {\bibinfo {title} {{Testing General Relativity with Low-Frequency, Space-Based Gravitational-Wave Detectors}},\ }\href {https://doi.org/10.12942/lrr-2013-7} {\bibfield  {journal} {\bibinfo  {journal} {Living Rev. Rel.}\ }\textbf {\bibinfo {volume} {16}},\ \bibinfo {pages} {7} (\bibinfo {year} {2013})},\ \Eprint {https://arxiv.org/abs/1212.5575} {arXiv:1212.5575 [gr-qc]} \BibitemShut {NoStop}%
\bibitem [{\citenamefont {Yunes}\ and\ \citenamefont {Pretorius}(2009)}]{ppE1}%
  \BibitemOpen
  \bibfield  {author} {\bibinfo {author} {\bibfnamefont {N.}~\bibnamefont {Yunes}}\ and\ \bibinfo {author} {\bibfnamefont {F.}~\bibnamefont {Pretorius}},\ }\bibfield  {title} {\bibinfo {title} {{Fundamental Theoretical Bias in Gravitational Wave Astrophysics and the Parameterized Post-Einsteinian Framework}},\ }\href {https://doi.org/10.1103/PhysRevD.80.122003} {\bibfield  {journal} {\bibinfo  {journal} {Phys. Rev. D}\ }\textbf {\bibinfo {volume} {80}},\ \bibinfo {pages} {122003} (\bibinfo {year} {2009})},\ \Eprint {https://arxiv.org/abs/0909.3328} {arXiv:0909.3328 [gr-qc]} \BibitemShut {NoStop}%
\bibitem [{\citenamefont {Liu}\ \emph {et~al.}(2020)\citenamefont {Liu}, \citenamefont {Ruan},\ and\ \citenamefont {Guo}}]{ppE_Taiji}%
  \BibitemOpen
  \bibfield  {author} {\bibinfo {author} {\bibfnamefont {C.}~\bibnamefont {Liu}}, \bibinfo {author} {\bibfnamefont {W.-H.}\ \bibnamefont {Ruan}},\ and\ \bibinfo {author} {\bibfnamefont {Z.-K.}\ \bibnamefont {Guo}},\ }\bibfield  {title} {\bibinfo {title} {{Constraining gravitational-wave polarizations with Taiji}},\ }\href {https://doi.org/10.1103/PhysRevD.102.124050} {\bibfield  {journal} {\bibinfo  {journal} {Phys. Rev. D}\ }\textbf {\bibinfo {volume} {102}},\ \bibinfo {pages} {124050} (\bibinfo {year} {2020})},\ \Eprint {https://arxiv.org/abs/2006.04413} {arXiv:2006.04413 [gr-qc]} \BibitemShut {NoStop}%
\bibitem [{\citenamefont {Chatziioannou}\ \emph {et~al.}(2012)\citenamefont {Chatziioannou}, \citenamefont {Yunes},\ and\ \citenamefont {Cornish}}]{ppE2}%
  \BibitemOpen
  \bibfield  {author} {\bibinfo {author} {\bibfnamefont {K.}~\bibnamefont {Chatziioannou}}, \bibinfo {author} {\bibfnamefont {N.}~\bibnamefont {Yunes}},\ and\ \bibinfo {author} {\bibfnamefont {N.}~\bibnamefont {Cornish}},\ }\bibfield  {title} {\bibinfo {title} {{Model-Independent Test of General Relativity: An Extended post-Einsteinian Framework with Complete Polarization Content}},\ }\href {https://doi.org/10.1103/PhysRevD.86.022004} {\bibfield  {journal} {\bibinfo  {journal} {Phys. Rev. D}\ }\textbf {\bibinfo {volume} {86}},\ \bibinfo {pages} {022004} (\bibinfo {year} {2012})},\ \bibinfo {note} {[Erratum: Phys.Rev.D 95, 129901 (2017)]},\ \Eprint {https://arxiv.org/abs/1204.2585} {arXiv:1204.2585 [gr-qc]} \BibitemShut {NoStop}%
\bibitem [{\citenamefont {Tahura}\ and\ \citenamefont {Yagi}(2018)}]{ppE3}%
  \BibitemOpen
  \bibfield  {author} {\bibinfo {author} {\bibfnamefont {S.}~\bibnamefont {Tahura}}\ and\ \bibinfo {author} {\bibfnamefont {K.}~\bibnamefont {Yagi}},\ }\bibfield  {title} {\bibinfo {title} {{Parameterized Post-Einsteinian Gravitational Waveforms in Various Modified Theories of Gravity}},\ }\href {https://doi.org/10.1103/PhysRevD.98.084042} {\bibfield  {journal} {\bibinfo  {journal} {Phys. Rev. D}\ }\textbf {\bibinfo {volume} {98}},\ \bibinfo {pages} {084042} (\bibinfo {year} {2018})},\ \bibinfo {note} {[Erratum: Phys.Rev.D 101, 109902 (2020)]},\ \Eprint {https://arxiv.org/abs/1809.00259} {arXiv:1809.00259 [gr-qc]} \BibitemShut {NoStop}%
\bibitem [{\citenamefont {Yunes}\ and\ \citenamefont {Siemens}(2013)}]{test_GR_PTA}%
  \BibitemOpen
  \bibfield  {author} {\bibinfo {author} {\bibfnamefont {N.}~\bibnamefont {Yunes}}\ and\ \bibinfo {author} {\bibfnamefont {X.}~\bibnamefont {Siemens}},\ }\bibfield  {title} {\bibinfo {title} {{Gravitational-Wave Tests of General Relativity with Ground-Based Detectors and Pulsar Timing-Arrays}},\ }\href {https://doi.org/10.12942/lrr-2013-9} {\bibfield  {journal} {\bibinfo  {journal} {Living Rev. Rel.}\ }\textbf {\bibinfo {volume} {16}},\ \bibinfo {pages} {9} (\bibinfo {year} {2013})},\ \Eprint {https://arxiv.org/abs/1304.3473} {arXiv:1304.3473 [gr-qc]} \BibitemShut {NoStop}%
\bibitem [{\citenamefont {Narikawa}\ and\ \citenamefont {Tagoshi}(2016)}]{ppE_ground}%
  \BibitemOpen
  \bibfield  {author} {\bibinfo {author} {\bibfnamefont {T.}~\bibnamefont {Narikawa}}\ and\ \bibinfo {author} {\bibfnamefont {H.}~\bibnamefont {Tagoshi}},\ }\bibfield  {title} {\bibinfo {title} {{The potential of advanced ground-based gravitational wave detectors to detect generic deviations from general relativity}},\ }\href {https://doi.org/10.1093/ptep/ptw126} {\bibfield  {journal} {\bibinfo  {journal} {PTEP}\ }\textbf {\bibinfo {volume} {2016}},\ \bibinfo {pages} {093E02} (\bibinfo {year} {2016})},\ \Eprint {https://arxiv.org/abs/1601.07691} {arXiv:1601.07691 [gr-qc]} \BibitemShut {NoStop}%
\bibitem [{\citenamefont {Takeda}\ \emph {et~al.}(2018)\citenamefont {Takeda}, \citenamefont {Nishizawa}, \citenamefont {Michimura}, \citenamefont {Nagano}, \citenamefont {Komori}, \citenamefont {Ando},\ and\ \citenamefont {Hayama}}]{CBC_FIM1}%
  \BibitemOpen
  \bibfield  {author} {\bibinfo {author} {\bibfnamefont {H.}~\bibnamefont {Takeda}}, \bibinfo {author} {\bibfnamefont {A.}~\bibnamefont {Nishizawa}}, \bibinfo {author} {\bibfnamefont {Y.}~\bibnamefont {Michimura}}, \bibinfo {author} {\bibfnamefont {K.}~\bibnamefont {Nagano}}, \bibinfo {author} {\bibfnamefont {K.}~\bibnamefont {Komori}}, \bibinfo {author} {\bibfnamefont {M.}~\bibnamefont {Ando}},\ and\ \bibinfo {author} {\bibfnamefont {K.}~\bibnamefont {Hayama}},\ }\bibfield  {title} {\bibinfo {title} {{Polarization test of gravitational waves from compact binary coalescences}},\ }\href {https://doi.org/10.1103/PhysRevD.98.022008} {\bibfield  {journal} {\bibinfo  {journal} {Phys. Rev. D}\ }\textbf {\bibinfo {volume} {98}},\ \bibinfo {pages} {022008} (\bibinfo {year} {2018})},\ \Eprint {https://arxiv.org/abs/1806.02182} {arXiv:1806.02182 [gr-qc]} \BibitemShut {NoStop}%
\bibitem [{\citenamefont {Takeda}\ \emph {et~al.}(2019)\citenamefont {Takeda}, \citenamefont {Nishizawa}, \citenamefont {Nagano}, \citenamefont {Michimura}, \citenamefont {Komori}, \citenamefont {Ando},\ and\ \citenamefont {Hayama}}]{CBC_FIM2}%
  \BibitemOpen
  \bibfield  {author} {\bibinfo {author} {\bibfnamefont {H.}~\bibnamefont {Takeda}}, \bibinfo {author} {\bibfnamefont {A.}~\bibnamefont {Nishizawa}}, \bibinfo {author} {\bibfnamefont {K.}~\bibnamefont {Nagano}}, \bibinfo {author} {\bibfnamefont {Y.}~\bibnamefont {Michimura}}, \bibinfo {author} {\bibfnamefont {K.}~\bibnamefont {Komori}}, \bibinfo {author} {\bibfnamefont {M.}~\bibnamefont {Ando}},\ and\ \bibinfo {author} {\bibfnamefont {K.}~\bibnamefont {Hayama}},\ }\bibfield  {title} {\bibinfo {title} {{Prospects for gravitational-wave polarization tests from compact binary mergers with future ground-based detectors}},\ }\href {https://doi.org/10.1103/PhysRevD.100.042001} {\bibfield  {journal} {\bibinfo  {journal} {Phys. Rev. D}\ }\textbf {\bibinfo {volume} {100}},\ \bibinfo {pages} {042001} (\bibinfo {year} {2019})},\ \Eprint {https://arxiv.org/abs/1904.09989} {arXiv:1904.09989 [gr-qc]} \BibitemShut {NoStop}%
\bibitem [{\citenamefont {Huwyler}\ \emph {et~al.}(2015)\citenamefont {Huwyler}, \citenamefont {Porter},\ and\ \citenamefont {Jetzer}}]{ppE_LISA}%
  \BibitemOpen
  \bibfield  {author} {\bibinfo {author} {\bibfnamefont {C.}~\bibnamefont {Huwyler}}, \bibinfo {author} {\bibfnamefont {E.~K.}\ \bibnamefont {Porter}},\ and\ \bibinfo {author} {\bibfnamefont {P.}~\bibnamefont {Jetzer}},\ }\bibfield  {title} {\bibinfo {title} {{Supermassive Black Hole Tests of General Relativity with eLISA}},\ }\href {https://doi.org/10.1103/PhysRevD.91.024037} {\bibfield  {journal} {\bibinfo  {journal} {Phys. Rev. D}\ }\textbf {\bibinfo {volume} {91}},\ \bibinfo {pages} {024037} (\bibinfo {year} {2015})},\ \Eprint {https://arxiv.org/abs/1410.8815} {arXiv:1410.8815 [gr-qc]} \BibitemShut {NoStop}%
\bibitem [{\citenamefont {Nair}\ \emph {et~al.}(2016)\citenamefont {Nair}, \citenamefont {Jhingan},\ and\ \citenamefont {Tanaka}}]{ppE_ET_DECIGO}%
  \BibitemOpen
  \bibfield  {author} {\bibinfo {author} {\bibfnamefont {R.}~\bibnamefont {Nair}}, \bibinfo {author} {\bibfnamefont {S.}~\bibnamefont {Jhingan}},\ and\ \bibinfo {author} {\bibfnamefont {T.}~\bibnamefont {Tanaka}},\ }\bibfield  {title} {\bibinfo {title} {{Synergy between ground and space based gravitational wave detectors for estimation of binary coalescence parameters}},\ }\href {https://doi.org/10.1093/ptep/ptw043} {\bibfield  {journal} {\bibinfo  {journal} {PTEP}\ }\textbf {\bibinfo {volume} {2016}},\ \bibinfo {pages} {053E01} (\bibinfo {year} {2016})},\ \Eprint {https://arxiv.org/abs/1504.04108} {arXiv:1504.04108 [gr-qc]} \BibitemShut {NoStop}%
\bibitem [{\citenamefont {Xie}\ \emph {et~al.}(2022)\citenamefont {Xie}, \citenamefont {Zhang}, \citenamefont {Huang}, \citenamefont {Hu},\ and\ \citenamefont {Mei}}]{DWD}%
  \BibitemOpen
  \bibfield  {author} {\bibinfo {author} {\bibfnamefont {N.}~\bibnamefont {Xie}}, \bibinfo {author} {\bibfnamefont {J.-d.}\ \bibnamefont {Zhang}}, \bibinfo {author} {\bibfnamefont {S.-J.}\ \bibnamefont {Huang}}, \bibinfo {author} {\bibfnamefont {Y.-M.}\ \bibnamefont {Hu}},\ and\ \bibinfo {author} {\bibfnamefont {J.}~\bibnamefont {Mei}},\ }\bibfield  {title} {\bibinfo {title} {{Constraining the extra polarization modes of gravitational waves with double white dwarfs}},\ }\href {https://doi.org/10.1103/PhysRevD.106.124017} {\bibfield  {journal} {\bibinfo  {journal} {Phys. Rev. D}\ }\textbf {\bibinfo {volume} {106}},\ \bibinfo {pages} {124017} (\bibinfo {year} {2022})},\ \Eprint {https://arxiv.org/abs/2208.10831} {arXiv:2208.10831 [gr-qc]} \BibitemShut {NoStop}%
\bibitem [{\citenamefont {Wang}\ and\ \citenamefont {Han}(2021{\natexlab{a}})}]{ppE_LISA_Taiji}%
  \BibitemOpen
  \bibfield  {author} {\bibinfo {author} {\bibfnamefont {G.}~\bibnamefont {Wang}}\ and\ \bibinfo {author} {\bibfnamefont {W.-B.}\ \bibnamefont {Han}},\ }\bibfield  {title} {\bibinfo {title} {{Observing gravitational wave polarizations with the LISA-TAIJI network}},\ }\href {https://doi.org/10.1103/PhysRevD.103.064021} {\bibfield  {journal} {\bibinfo  {journal} {Phys. Rev. D}\ }\textbf {\bibinfo {volume} {103}},\ \bibinfo {pages} {064021} (\bibinfo {year} {2021}{\natexlab{a}})},\ \Eprint {https://arxiv.org/abs/2101.01991} {arXiv:2101.01991 [gr-qc]} \BibitemShut {NoStop}%
\bibitem [{\citenamefont {Wu}\ and\ \citenamefont {Li}(2023)}]{my_paper2}%
  \BibitemOpen
  \bibfield  {author} {\bibinfo {author} {\bibfnamefont {J.}~\bibnamefont {Wu}}\ and\ \bibinfo {author} {\bibfnamefont {J.}~\bibnamefont {Li}},\ }\bibfield  {title} {\bibinfo {title} {{Subtraction of the confusion foreground and parameter uncertainty of resolvable galactic binaries on the networks of space-based gravitational-wave detectors}},\ }\href {https://doi.org/10.1103/PhysRevD.108.124047} {\bibfield  {journal} {\bibinfo  {journal} {Phys. Rev. D}\ }\textbf {\bibinfo {volume} {108}},\ \bibinfo {pages} {124047} (\bibinfo {year} {2023})},\ \Eprint {https://arxiv.org/abs/2307.05568} {arXiv:2307.05568 [gr-qc]} \BibitemShut {NoStop}%
\bibitem [{\citenamefont {Wu}\ \emph {et~al.}(2024)\citenamefont {Wu}, \citenamefont {Li}, \citenamefont {Liu},\ and\ \citenamefont {Cao}}]{my_paper3}%
  \BibitemOpen
  \bibfield  {author} {\bibinfo {author} {\bibfnamefont {J.}~\bibnamefont {Wu}}, \bibinfo {author} {\bibfnamefont {J.}~\bibnamefont {Li}}, \bibinfo {author} {\bibfnamefont {X.}~\bibnamefont {Liu}},\ and\ \bibinfo {author} {\bibfnamefont {Z.}~\bibnamefont {Cao}},\ }\bibfield  {title} {\bibinfo {title} {{Comparison and application of different post-Newtonian models for inspiralling stellar-mass binary black holes with space-based GW detectors}},\ }\href {https://doi.org/10.1103/PhysRevD.109.104014} {\bibfield  {journal} {\bibinfo  {journal} {Phys. Rev. D}\ }\textbf {\bibinfo {volume} {109}},\ \bibinfo {pages} {104014} (\bibinfo {year} {2024})},\ \Eprint {https://arxiv.org/abs/2401.03113} {arXiv:2401.03113 [gr-qc]} \BibitemShut {NoStop}%
\bibitem [{\citenamefont {Wu}\ \emph {et~al.}(2023)\citenamefont {Wu}, \citenamefont {Li},\ and\ \citenamefont {Jiang}}]{my_paper1}%
  \BibitemOpen
  \bibfield  {author} {\bibinfo {author} {\bibfnamefont {J.}~\bibnamefont {Wu}}, \bibinfo {author} {\bibfnamefont {J.}~\bibnamefont {Li}},\ and\ \bibinfo {author} {\bibfnamefont {Q.-Q.}\ \bibnamefont {Jiang}},\ }\bibfield  {title} {\bibinfo {title} {{Application of Newtonian approximate model to LIGO gravitational wave data processing}},\ }\href {https://doi.org/10.1088/1674-1056/acd8a3} {\bibfield  {journal} {\bibinfo  {journal} {Chin. Phys. B}\ }\textbf {\bibinfo {volume} {32}},\ \bibinfo {pages} {090401} (\bibinfo {year} {2023})}\BibitemShut {NoStop}%
\bibitem [{\citenamefont {Hansen}\ \emph {et~al.}(2015)\citenamefont {Hansen}, \citenamefont {Yunes},\ and\ \citenamefont {Yagi}}]{ppE4}%
  \BibitemOpen
  \bibfield  {author} {\bibinfo {author} {\bibfnamefont {D.}~\bibnamefont {Hansen}}, \bibinfo {author} {\bibfnamefont {N.}~\bibnamefont {Yunes}},\ and\ \bibinfo {author} {\bibfnamefont {K.}~\bibnamefont {Yagi}},\ }\bibfield  {title} {\bibinfo {title} {{Projected Constraints on Lorentz-Violating Gravity with Gravitational Waves}},\ }\href {https://doi.org/10.1103/PhysRevD.91.082003} {\bibfield  {journal} {\bibinfo  {journal} {Phys. Rev. D}\ }\textbf {\bibinfo {volume} {91}},\ \bibinfo {pages} {082003} (\bibinfo {year} {2015})},\ \Eprint {https://arxiv.org/abs/1412.4132} {arXiv:1412.4132 [gr-qc]} \BibitemShut {NoStop}%
\bibitem [{\citenamefont {Arun}(2012)}]{dipolar}%
  \BibitemOpen
  \bibfield  {author} {\bibinfo {author} {\bibfnamefont {K.~G.}\ \bibnamefont {Arun}},\ }\bibfield  {title} {\bibinfo {title} {{Generic bounds on dipolar gravitational radiation from inspiralling compact binaries}},\ }\href {https://doi.org/10.1088/0264-9381/29/7/075011} {\bibfield  {journal} {\bibinfo  {journal} {Class. Quant. Grav.}\ }\textbf {\bibinfo {volume} {29}},\ \bibinfo {pages} {075011} (\bibinfo {year} {2012})},\ \Eprint {https://arxiv.org/abs/1202.5911} {arXiv:1202.5911 [gr-qc]} \BibitemShut {NoStop}%
\bibitem [{\citenamefont {O'Beirne}\ \emph {et~al.}(2019)\citenamefont {O'Beirne}, \citenamefont {Cornish}, \citenamefont {Vigeland},\ and\ \citenamefont {Taylor}}]{ppE_PTA}%
  \BibitemOpen
  \bibfield  {author} {\bibinfo {author} {\bibfnamefont {L.}~\bibnamefont {O'Beirne}}, \bibinfo {author} {\bibfnamefont {N.~J.}\ \bibnamefont {Cornish}}, \bibinfo {author} {\bibfnamefont {S.~J.}\ \bibnamefont {Vigeland}},\ and\ \bibinfo {author} {\bibfnamefont {S.~R.}\ \bibnamefont {Taylor}},\ }\bibfield  {title} {\bibinfo {title} {{Constraining alternative polarization states of gravitational waves from individual black hole binaries using pulsar timing arrays}},\ }\href {https://doi.org/10.1103/PhysRevD.99.124039} {\bibfield  {journal} {\bibinfo  {journal} {Phys. Rev. D}\ }\textbf {\bibinfo {volume} {99}},\ \bibinfo {pages} {124039} (\bibinfo {year} {2019})},\ \Eprint {https://arxiv.org/abs/1904.02744} {arXiv:1904.02744 [gr-qc]} \BibitemShut {NoStop}%
\bibitem [{\citenamefont {Wang}\ \emph {et~al.}(2021)\citenamefont {Wang}, \citenamefont {Ni}, \citenamefont {Han}, \citenamefont {Xu},\ and\ \citenamefont {Luo}}]{Alternative_LISA_TAIJI1}%
  \BibitemOpen
  \bibfield  {author} {\bibinfo {author} {\bibfnamefont {G.}~\bibnamefont {Wang}}, \bibinfo {author} {\bibfnamefont {W.-T.}\ \bibnamefont {Ni}}, \bibinfo {author} {\bibfnamefont {W.-B.}\ \bibnamefont {Han}}, \bibinfo {author} {\bibfnamefont {P.}~\bibnamefont {Xu}},\ and\ \bibinfo {author} {\bibfnamefont {Z.}~\bibnamefont {Luo}},\ }\bibfield  {title} {\bibinfo {title} {{Alternative LISA-TAIJI networks}},\ }\href {https://doi.org/10.1103/PhysRevD.104.024012} {\bibfield  {journal} {\bibinfo  {journal} {Phys. Rev. D}\ }\textbf {\bibinfo {volume} {104}},\ \bibinfo {pages} {024012} (\bibinfo {year} {2021})},\ \Eprint {https://arxiv.org/abs/2105.00746} {arXiv:2105.00746 [gr-qc]} \BibitemShut {NoStop}%
\bibitem [{\citenamefont {Wang}\ and\ \citenamefont {Han}(2021{\natexlab{b}})}]{Alternative_LISA_TAIJI2}%
  \BibitemOpen
  \bibfield  {author} {\bibinfo {author} {\bibfnamefont {G.}~\bibnamefont {Wang}}\ and\ \bibinfo {author} {\bibfnamefont {W.-B.}\ \bibnamefont {Han}},\ }\bibfield  {title} {\bibinfo {title} {{Alternative LISA-TAIJI networks: Detectability of the isotropic stochastic gravitational wave background}},\ }\href {https://doi.org/10.1103/PhysRevD.104.104015} {\bibfield  {journal} {\bibinfo  {journal} {Phys. Rev. D}\ }\textbf {\bibinfo {volume} {104}},\ \bibinfo {pages} {104015} (\bibinfo {year} {2021}{\natexlab{b}})},\ \Eprint {https://arxiv.org/abs/2108.11151} {arXiv:2108.11151 [gr-qc]} \BibitemShut {NoStop}%
\bibitem [{\citenamefont {Huang}\ \emph {et~al.}(2020{\natexlab{a}})\citenamefont {Huang}, \citenamefont {Hu}, \citenamefont {Korol}, \citenamefont {Li}, \citenamefont {Liang}, \citenamefont {Lu}, \citenamefont {Wang}, \citenamefont {Yu},\ and\ \citenamefont {Mei}}]{Alternative_TianQin}%
  \BibitemOpen
  \bibfield  {author} {\bibinfo {author} {\bibfnamefont {S.-J.}\ \bibnamefont {Huang}}, \bibinfo {author} {\bibfnamefont {Y.-M.}\ \bibnamefont {Hu}}, \bibinfo {author} {\bibfnamefont {V.}~\bibnamefont {Korol}}, \bibinfo {author} {\bibfnamefont {P.-C.}\ \bibnamefont {Li}}, \bibinfo {author} {\bibfnamefont {Z.-C.}\ \bibnamefont {Liang}}, \bibinfo {author} {\bibfnamefont {Y.}~\bibnamefont {Lu}}, \bibinfo {author} {\bibfnamefont {H.-T.}\ \bibnamefont {Wang}}, \bibinfo {author} {\bibfnamefont {S.}~\bibnamefont {Yu}},\ and\ \bibinfo {author} {\bibfnamefont {J.}~\bibnamefont {Mei}},\ }\bibfield  {title} {\bibinfo {title} {{Science with the TianQin Observatory: Preliminary results on Galactic double white dwarf binaries}},\ }\href {https://doi.org/10.1103/PhysRevD.102.063021} {\bibfield  {journal} {\bibinfo  {journal} {Phys. Rev. D}\ }\textbf {\bibinfo {volume} {102}},\ \bibinfo {pages} {063021} (\bibinfo {year} {2020}{\natexlab{a}})},\ \Eprint {https://arxiv.org/abs/2005.07889} {arXiv:2005.07889 [astro-ph.HE]}
  \BibitemShut {NoStop}%
\bibitem [{\citenamefont {Cai}\ \emph {et~al.}(2023)\citenamefont {Cai}, \citenamefont {Guo}, \citenamefont {Hu}, \citenamefont {Liu}, \citenamefont {Lu}, \citenamefont {Ni}, \citenamefont {Ruan}, \citenamefont {Seto}, \citenamefont {Wang},\ and\ \citenamefont {Wu}}]{space_network}%
  \BibitemOpen
  \bibfield  {author} {\bibinfo {author} {\bibfnamefont {R.-G.}\ \bibnamefont {Cai}}, \bibinfo {author} {\bibfnamefont {Z.-K.}\ \bibnamefont {Guo}}, \bibinfo {author} {\bibfnamefont {B.}~\bibnamefont {Hu}}, \bibinfo {author} {\bibfnamefont {C.}~\bibnamefont {Liu}}, \bibinfo {author} {\bibfnamefont {Y.}~\bibnamefont {Lu}}, \bibinfo {author} {\bibfnamefont {W.-T.}\ \bibnamefont {Ni}}, \bibinfo {author} {\bibfnamefont {W.-H.}\ \bibnamefont {Ruan}}, \bibinfo {author} {\bibfnamefont {N.}~\bibnamefont {Seto}}, \bibinfo {author} {\bibfnamefont {G.}~\bibnamefont {Wang}},\ and\ \bibinfo {author} {\bibfnamefont {Y.-L.}\ \bibnamefont {Wu}},\ }\href {https://arxiv.org/abs/2305.04551} {\bibinfo {title} {On networks of space-based gravitational-wave detectors}} (\bibinfo {year} {2023}),\ \Eprint {https://arxiv.org/abs/2305.04551} {arXiv:2305.04551 [gr-qc]} \BibitemShut {NoStop}%
\bibitem [{\citenamefont {{LIGO Scientific Collaboration}}\ \emph {et~al.}(2018)\citenamefont {{LIGO Scientific Collaboration}}, \citenamefont {{Virgo Collaboration}},\ and\ \citenamefont {{KAGRA Collaboration}}}]{lalsuite1}%
  \BibitemOpen
  \bibfield  {author} {\bibinfo {author} {\bibnamefont {{LIGO Scientific Collaboration}}}, \bibinfo {author} {\bibnamefont {{Virgo Collaboration}}},\ and\ \bibinfo {author} {\bibnamefont {{KAGRA Collaboration}}},\ }\href {https://doi.org/10.7935/GT1W-FZ16} {\bibinfo {title} {{LVK} {A}lgorithm {L}ibrary - {LALS}uite}},\ \bibinfo {howpublished} {Free software (GPL)} (\bibinfo {year} {2018})\BibitemShut {NoStop}%
\bibitem [{\citenamefont {Wette}(2020)}]{lalsuite2}%
  \BibitemOpen
  \bibfield  {author} {\bibinfo {author} {\bibfnamefont {K.}~\bibnamefont {Wette}},\ }\bibfield  {title} {\bibinfo {title} {{SWIGLAL: Python and Octave interfaces to the LALSuite gravitational-wave data analysis libraries}},\ }\href {https://doi.org/10.1016/j.softx.2020.100634} {\bibfield  {journal} {\bibinfo  {journal} {SoftwareX}\ }\textbf {\bibinfo {volume} {12}},\ \bibinfo {pages} {100634} (\bibinfo {year} {2020})}\BibitemShut {NoStop}%
\bibitem [{\citenamefont {Murdin}(2001)}]{obliquity_of_ecliptic}%
  \BibitemOpen
  \bibinfo {editor} {\bibfnamefont {P.}~\bibnamefont {Murdin}},\ ed.,\ \href@noop {} {\emph {\bibinfo {title} {Encyclopedia of Astronomy \& Astrophysics}}},\ \bibinfo {edition} {1st}\ ed.\ (\bibinfo  {publisher} {CRC Press},\ \bibinfo {year} {2001})\ \bibinfo {note} {\url{https://doi.org/10.1201/9781003220435}}\BibitemShut {NoStop}%
\bibitem [{\citenamefont {Robson}\ \emph {et~al.}(2019)\citenamefont {Robson}, \citenamefont {Cornish},\ and\ \citenamefont {Liu}}]{LISA_noise}%
  \BibitemOpen
  \bibfield  {author} {\bibinfo {author} {\bibfnamefont {T.}~\bibnamefont {Robson}}, \bibinfo {author} {\bibfnamefont {N.~J.}\ \bibnamefont {Cornish}},\ and\ \bibinfo {author} {\bibfnamefont {C.}~\bibnamefont {Liu}},\ }\bibfield  {title} {\bibinfo {title} {{The construction and use of LISA sensitivity curves}},\ }\href {https://doi.org/10.1088/1361-6382/ab1101} {\bibfield  {journal} {\bibinfo  {journal} {Class. Quant. Grav.}\ }\textbf {\bibinfo {volume} {36}},\ \bibinfo {pages} {105011} (\bibinfo {year} {2019})},\ \Eprint {https://arxiv.org/abs/1803.01944} {arXiv:1803.01944 [astro-ph.HE]} \BibitemShut {NoStop}%
\bibitem [{\citenamefont {Ren}\ \emph {et~al.}(2023)\citenamefont {Ren}, \citenamefont {Zhao}, \citenamefont {Cao}, \citenamefont {Guo}, \citenamefont {Han}, \citenamefont {Jin},\ and\ \citenamefont {Wu}}]{Taiji_noise}%
  \BibitemOpen
  \bibfield  {author} {\bibinfo {author} {\bibfnamefont {Z.}~\bibnamefont {Ren}}, \bibinfo {author} {\bibfnamefont {T.}~\bibnamefont {Zhao}}, \bibinfo {author} {\bibfnamefont {Z.}~\bibnamefont {Cao}}, \bibinfo {author} {\bibfnamefont {Z.-K.}\ \bibnamefont {Guo}}, \bibinfo {author} {\bibfnamefont {W.-B.}\ \bibnamefont {Han}}, \bibinfo {author} {\bibfnamefont {H.-B.}\ \bibnamefont {Jin}},\ and\ \bibinfo {author} {\bibfnamefont {Y.-L.}\ \bibnamefont {Wu}},\ }\bibfield  {title} {\bibinfo {title} {{Taiji data challenge for exploring gravitational wave universe}},\ }\href {https://doi.org/10.1007/s11467-023-1318-y} {\bibfield  {journal} {\bibinfo  {journal} {Front. Phys. (Beijing)}\ }\textbf {\bibinfo {volume} {18}},\ \bibinfo {pages} {64302} (\bibinfo {year} {2023})},\ \Eprint {https://arxiv.org/abs/2301.02967} {arXiv:2301.02967 [gr-qc]} \BibitemShut {NoStop}%
\bibitem [{\citenamefont {Hu}\ \emph {et~al.}(2018)\citenamefont {Hu}, \citenamefont {Li}, \citenamefont {Wang}, \citenamefont {Feng}, \citenamefont {Zhou}, \citenamefont {Hu}, \citenamefont {Hu}, \citenamefont {Mei},\ and\ \citenamefont {Shao}}]{TianQin_noise}%
  \BibitemOpen
  \bibfield  {author} {\bibinfo {author} {\bibfnamefont {X.-C.}\ \bibnamefont {Hu}}, \bibinfo {author} {\bibfnamefont {X.-H.}\ \bibnamefont {Li}}, \bibinfo {author} {\bibfnamefont {Y.}~\bibnamefont {Wang}}, \bibinfo {author} {\bibfnamefont {W.-F.}\ \bibnamefont {Feng}}, \bibinfo {author} {\bibfnamefont {M.-Y.}\ \bibnamefont {Zhou}}, \bibinfo {author} {\bibfnamefont {Y.-M.}\ \bibnamefont {Hu}}, \bibinfo {author} {\bibfnamefont {S.-C.}\ \bibnamefont {Hu}}, \bibinfo {author} {\bibfnamefont {J.-W.}\ \bibnamefont {Mei}},\ and\ \bibinfo {author} {\bibfnamefont {C.-G.}\ \bibnamefont {Shao}},\ }\bibfield  {title} {\bibinfo {title} {{Fundamentals of the orbit and response for TianQin}},\ }\href {https://doi.org/10.1088/1361-6382/aab52f} {\bibfield  {journal} {\bibinfo  {journal} {Class. Quant. Grav.}\ }\textbf {\bibinfo {volume} {35}},\ \bibinfo {pages} {095008} (\bibinfo {year} {2018})},\ \Eprint {https://arxiv.org/abs/1803.03368} {arXiv:1803.03368 [gr-qc]} \BibitemShut {NoStop}%
\bibitem [{\citenamefont {Punturo}\ \emph {et~al.}(2010)\citenamefont {Punturo} \emph {et~al.}}]{ET}%
  \BibitemOpen
  \bibfield  {author} {\bibinfo {author} {\bibfnamefont {M.}~\bibnamefont {Punturo}} \emph {et~al.},\ }\bibfield  {title} {\bibinfo {title} {{The Einstein Telescope: A third-generation gravitational wave observatory}},\ }\href {https://doi.org/10.1088/0264-9381/27/19/194002} {\bibfield  {journal} {\bibinfo  {journal} {Class. Quant. Grav.}\ }\textbf {\bibinfo {volume} {27}},\ \bibinfo {pages} {194002} (\bibinfo {year} {2010})}\BibitemShut {NoStop}%
\bibitem [{\citenamefont {Yi}\ \emph {et~al.}(2022)\citenamefont {Yi}, \citenamefont {Nelemans}, \citenamefont {Brinkerink}, \citenamefont {Kostrzewa-Rutkowska}, \citenamefont {Timmer}, \citenamefont {Stoppa}, \citenamefont {Rossi},\ and\ \citenamefont {Portegies~Zwart}}]{detector_noise}%
  \BibitemOpen
  \bibfield  {author} {\bibinfo {author} {\bibfnamefont {S.-X.}\ \bibnamefont {Yi}}, \bibinfo {author} {\bibfnamefont {G.}~\bibnamefont {Nelemans}}, \bibinfo {author} {\bibfnamefont {C.}~\bibnamefont {Brinkerink}}, \bibinfo {author} {\bibfnamefont {Z.}~\bibnamefont {Kostrzewa-Rutkowska}}, \bibinfo {author} {\bibfnamefont {S.~T.}\ \bibnamefont {Timmer}}, \bibinfo {author} {\bibfnamefont {F.}~\bibnamefont {Stoppa}}, \bibinfo {author} {\bibfnamefont {E.~M.}\ \bibnamefont {Rossi}},\ and\ \bibinfo {author} {\bibfnamefont {S.~F.}\ \bibnamefont {Portegies~Zwart}},\ }\bibfield  {title} {\bibinfo {title} {{The Gravitational Wave Universe Toolbox - A software package to simulate observations of the gravitational wave universe with different detectors,}},\ }\href {https://doi.org/10.1051/0004-6361/202141634} {\bibfield  {journal} {\bibinfo  {journal} {Astron. Astrophys.}\ }\textbf {\bibinfo {volume} {663}},\ \bibinfo {pages} {A155} (\bibinfo {year} {2022})},\ \Eprint {https://arxiv.org/abs/2106.13662} {arXiv:2106.13662
  [astro-ph.HE]} \BibitemShut {NoStop}%
\bibitem [{\citenamefont {Hild}\ \emph {et~al.}(2011)\citenamefont {Hild} \emph {et~al.}}]{ET_noise}%
  \BibitemOpen
  \bibfield  {author} {\bibinfo {author} {\bibfnamefont {S.}~\bibnamefont {Hild}} \emph {et~al.},\ }\bibfield  {title} {\bibinfo {title} {{Sensitivity Studies for Third-Generation Gravitational Wave Observatories}},\ }\href {https://doi.org/10.1088/0264-9381/28/9/094013} {\bibfield  {journal} {\bibinfo  {journal} {Class. Quant. Grav.}\ }\textbf {\bibinfo {volume} {28}},\ \bibinfo {pages} {094013} (\bibinfo {year} {2011})},\ \Eprint {https://arxiv.org/abs/1012.0908} {arXiv:1012.0908 [gr-qc]} \BibitemShut {NoStop}%
\bibitem [{\citenamefont {Nishizawa}\ \emph {et~al.}(2009)\citenamefont {Nishizawa}, \citenamefont {Taruya}, \citenamefont {Hayama}, \citenamefont {Kawamura},\ and\ \citenamefont {Sakagami}}]{polarization_tensor}%
  \BibitemOpen
  \bibfield  {author} {\bibinfo {author} {\bibfnamefont {A.}~\bibnamefont {Nishizawa}}, \bibinfo {author} {\bibfnamefont {A.}~\bibnamefont {Taruya}}, \bibinfo {author} {\bibfnamefont {K.}~\bibnamefont {Hayama}}, \bibinfo {author} {\bibfnamefont {S.}~\bibnamefont {Kawamura}},\ and\ \bibinfo {author} {\bibfnamefont {M.-a.}\ \bibnamefont {Sakagami}},\ }\bibfield  {title} {\bibinfo {title} {{Probing non-tensorial polarizations of stochastic gravitational-wave backgrounds with ground-based laser interferometers}},\ }\href {https://doi.org/10.1103/PhysRevD.79.082002} {\bibfield  {journal} {\bibinfo  {journal} {Phys. Rev. D}\ }\textbf {\bibinfo {volume} {79}},\ \bibinfo {pages} {082002} (\bibinfo {year} {2009})},\ \Eprint {https://arxiv.org/abs/0903.0528} {arXiv:0903.0528 [astro-ph.CO]} \BibitemShut {NoStop}%
\bibitem [{\citenamefont {Cornish}\ and\ \citenamefont {Larson}(2001)}]{transfer_function}%
  \BibitemOpen
  \bibfield  {author} {\bibinfo {author} {\bibfnamefont {N.~J.}\ \bibnamefont {Cornish}}\ and\ \bibinfo {author} {\bibfnamefont {S.~L.}\ \bibnamefont {Larson}},\ }\bibfield  {title} {\bibinfo {title} {{Space missions to detect the cosmic gravitational wave background}},\ }\href {https://doi.org/10.1088/0264-9381/18/17/308} {\bibfield  {journal} {\bibinfo  {journal} {Class. Quant. Grav.}\ }\textbf {\bibinfo {volume} {18}},\ \bibinfo {pages} {3473} (\bibinfo {year} {2001})},\ \Eprint {https://arxiv.org/abs/gr-qc/0103075} {arXiv:gr-qc/0103075} \BibitemShut {NoStop}%
\bibitem [{\citenamefont {Shah}\ \emph {et~al.}(2012)\citenamefont {Shah}, \citenamefont {van~der Sluys},\ and\ \citenamefont {Nelemans}}]{FIM1}%
  \BibitemOpen
  \bibfield  {author} {\bibinfo {author} {\bibfnamefont {S.}~\bibnamefont {Shah}}, \bibinfo {author} {\bibfnamefont {M.}~\bibnamefont {van~der Sluys}},\ and\ \bibinfo {author} {\bibfnamefont {G.}~\bibnamefont {Nelemans}},\ }\bibfield  {title} {\bibinfo {title} {{Using electromagnetic observations to aid gravitational-wave parameter estimation of compact binaries observed with LISA}},\ }\href {https://doi.org/10.1051/0004-6361/201219309} {\bibfield  {journal} {\bibinfo  {journal} {Astron. Astrophys.}\ }\textbf {\bibinfo {volume} {544}},\ \bibinfo {pages} {A153} (\bibinfo {year} {2012})},\ \Eprint {https://arxiv.org/abs/1207.6770} {arXiv:1207.6770 [astro-ph.IM]} \BibitemShut {NoStop}%
\bibitem [{\citenamefont {Vallisneri}(2008)}]{FIM2}%
  \BibitemOpen
  \bibfield  {author} {\bibinfo {author} {\bibfnamefont {M.}~\bibnamefont {Vallisneri}},\ }\bibfield  {title} {\bibinfo {title} {{Use and abuse of the Fisher information matrix in the assessment of gravitational-wave parameter-estimation prospects}},\ }\href {https://doi.org/10.1103/PhysRevD.77.042001} {\bibfield  {journal} {\bibinfo  {journal} {Phys. Rev. D}\ }\textbf {\bibinfo {volume} {77}},\ \bibinfo {pages} {042001} (\bibinfo {year} {2008})},\ \Eprint {https://arxiv.org/abs/gr-qc/0703086} {arXiv:gr-qc/0703086} \BibitemShut {NoStop}%
\bibitem [{\citenamefont {Huang}\ \emph {et~al.}(2020{\natexlab{b}})\citenamefont {Huang}, \citenamefont {Hu}, \citenamefont {Korol}, \citenamefont {Li}, \citenamefont {Liang}, \citenamefont {Lu}, \citenamefont {Wang}, \citenamefont {Yu},\ and\ \citenamefont {Mei}}]{TianQinGBs}%
  \BibitemOpen
  \bibfield  {author} {\bibinfo {author} {\bibfnamefont {S.-J.}\ \bibnamefont {Huang}}, \bibinfo {author} {\bibfnamefont {Y.-M.}\ \bibnamefont {Hu}}, \bibinfo {author} {\bibfnamefont {V.}~\bibnamefont {Korol}}, \bibinfo {author} {\bibfnamefont {P.-C.}\ \bibnamefont {Li}}, \bibinfo {author} {\bibfnamefont {Z.-C.}\ \bibnamefont {Liang}}, \bibinfo {author} {\bibfnamefont {Y.}~\bibnamefont {Lu}}, \bibinfo {author} {\bibfnamefont {H.-T.}\ \bibnamefont {Wang}}, \bibinfo {author} {\bibfnamefont {S.}~\bibnamefont {Yu}},\ and\ \bibinfo {author} {\bibfnamefont {J.}~\bibnamefont {Mei}},\ }\bibfield  {title} {\bibinfo {title} {{Science with the TianQin Observatory: Preliminary results on Galactic double white dwarf binaries}},\ }\href {https://doi.org/10.1103/PhysRevD.102.063021} {\bibfield  {journal} {\bibinfo  {journal} {Phys. Rev. D}\ }\textbf {\bibinfo {volume} {102}},\ \bibinfo {pages} {063021} (\bibinfo {year} {2020}{\natexlab{b}})},\ \Eprint {https://arxiv.org/abs/2005.07889} {arXiv:2005.07889 [astro-ph.HE]}
  \BibitemShut {NoStop}%
\bibitem [{\citenamefont {Cutler}(1998)}]{Omega}%
  \BibitemOpen
  \bibfield  {author} {\bibinfo {author} {\bibfnamefont {C.}~\bibnamefont {Cutler}},\ }\bibfield  {title} {\bibinfo {title} {{Angular resolution of the LISA gravitational wave detector}},\ }\href {https://doi.org/10.1103/PhysRevD.57.7089} {\bibfield  {journal} {\bibinfo  {journal} {Phys. Rev. D}\ }\textbf {\bibinfo {volume} {57}},\ \bibinfo {pages} {7089} (\bibinfo {year} {1998})},\ \Eprint {https://arxiv.org/abs/gr-qc/9703068} {arXiv:gr-qc/9703068} \BibitemShut {NoStop}%
\bibitem [{\citenamefont {Ewing}\ \emph {et~al.}(2021)\citenamefont {Ewing}, \citenamefont {Sachdev}, \citenamefont {Borhanian},\ and\ \citenamefont {Sathyaprakash}}]{LISA_SBBH}%
  \BibitemOpen
  \bibfield  {author} {\bibinfo {author} {\bibfnamefont {B.}~\bibnamefont {Ewing}}, \bibinfo {author} {\bibfnamefont {S.}~\bibnamefont {Sachdev}}, \bibinfo {author} {\bibfnamefont {S.}~\bibnamefont {Borhanian}},\ and\ \bibinfo {author} {\bibfnamefont {B.~S.}\ \bibnamefont {Sathyaprakash}},\ }\bibfield  {title} {\bibinfo {title} {{Archival searches for stellar-mass binary black holes in LISA data}},\ }\href {https://doi.org/10.1103/PhysRevD.103.023025} {\bibfield  {journal} {\bibinfo  {journal} {Phys. Rev. D}\ }\textbf {\bibinfo {volume} {103}},\ \bibinfo {pages} {023025} (\bibinfo {year} {2021})},\ \Eprint {https://arxiv.org/abs/2011.03036} {arXiv:2011.03036 [gr-qc]} \BibitemShut {NoStop}%
\bibitem [{\citenamefont {Guti\'errez}\ \emph {et~al.}(2022)\citenamefont {Guti\'errez}, \citenamefont {Combi}, \citenamefont {Noble}, \citenamefont {Campanelli}, \citenamefont {Krolik}, \citenamefont {Armengol},\ and\ \citenamefont {Garc\'\i{}a}}]{EM_MBHB1}%
  \BibitemOpen
  \bibfield  {author} {\bibinfo {author} {\bibfnamefont {E.~M.}\ \bibnamefont {Guti\'errez}}, \bibinfo {author} {\bibfnamefont {L.}~\bibnamefont {Combi}}, \bibinfo {author} {\bibfnamefont {S.~C.}\ \bibnamefont {Noble}}, \bibinfo {author} {\bibfnamefont {M.}~\bibnamefont {Campanelli}}, \bibinfo {author} {\bibfnamefont {J.~H.}\ \bibnamefont {Krolik}}, \bibinfo {author} {\bibfnamefont {F.~G.~L.}\ \bibnamefont {Armengol}},\ and\ \bibinfo {author} {\bibfnamefont {F.}~\bibnamefont {Garc\'\i{}a}},\ }\bibfield  {title} {\bibinfo {title} {{Electromagnetic Signatures from Supermassive Binary Black Holes Approaching Merger}},\ }\href {https://doi.org/10.3847/1538-4357/ac56de} {\bibfield  {journal} {\bibinfo  {journal} {Astrophys. J.}\ }\textbf {\bibinfo {volume} {928}},\ \bibinfo {pages} {137} (\bibinfo {year} {2022})},\ \Eprint {https://arxiv.org/abs/2112.09773} {arXiv:2112.09773 [astro-ph.HE]} \BibitemShut {NoStop}%
\bibitem [{\citenamefont {Roedig}\ \emph {et~al.}(2014)\citenamefont {Roedig}, \citenamefont {Krolik},\ and\ \citenamefont {Coleman~Miller}}]{EM_MBHB2}%
  \BibitemOpen
  \bibfield  {author} {\bibinfo {author} {\bibfnamefont {C.}~\bibnamefont {Roedig}}, \bibinfo {author} {\bibfnamefont {J.~H.}\ \bibnamefont {Krolik}},\ and\ \bibinfo {author} {\bibfnamefont {M.}~\bibnamefont {Coleman~Miller}},\ }\bibfield  {title} {\bibinfo {title} {{Observational Signatures of Binary Supermassive Black Holes}},\ }\href {https://doi.org/10.1088/0004-637X/785/2/115} {\bibfield  {journal} {\bibinfo  {journal} {Astrophys. J.}\ }\textbf {\bibinfo {volume} {785}},\ \bibinfo {pages} {115} (\bibinfo {year} {2014})},\ \Eprint {https://arxiv.org/abs/1402.7098} {arXiv:1402.7098 [astro-ph.HE]} \BibitemShut {NoStop}%
\bibitem [{\citenamefont {Tagawa}\ \emph {et~al.}(2023)\citenamefont {Tagawa}, \citenamefont {Kimura}, \citenamefont {Haiman}, \citenamefont {Perna},\ and\ \citenamefont {Bartos}}]{EM_SBBH1}%
  \BibitemOpen
  \bibfield  {author} {\bibinfo {author} {\bibfnamefont {H.}~\bibnamefont {Tagawa}}, \bibinfo {author} {\bibfnamefont {S.~S.}\ \bibnamefont {Kimura}}, \bibinfo {author} {\bibfnamefont {Z.}~\bibnamefont {Haiman}}, \bibinfo {author} {\bibfnamefont {R.}~\bibnamefont {Perna}},\ and\ \bibinfo {author} {\bibfnamefont {I.}~\bibnamefont {Bartos}},\ }\bibfield  {title} {\bibinfo {title} {{Observable Signature of Merging Stellar-mass Black Holes in Active Galactic Nuclei}},\ }\href {https://doi.org/10.3847/1538-4357/acc4bb} {\bibfield  {journal} {\bibinfo  {journal} {Astrophys. J.}\ }\textbf {\bibinfo {volume} {950}},\ \bibinfo {pages} {13} (\bibinfo {year} {2023})},\ \Eprint {https://arxiv.org/abs/2301.07111} {arXiv:2301.07111 [astro-ph.HE]} \BibitemShut {NoStop}%
\bibitem [{\citenamefont {Bartos}\ \emph {et~al.}(2017)\citenamefont {Bartos}, \citenamefont {Kocsis}, \citenamefont {Haiman},\ and\ \citenamefont {M\'arka}}]{EM_SBBH2}%
  \BibitemOpen
  \bibfield  {author} {\bibinfo {author} {\bibfnamefont {I.}~\bibnamefont {Bartos}}, \bibinfo {author} {\bibfnamefont {B.}~\bibnamefont {Kocsis}}, \bibinfo {author} {\bibfnamefont {Z.}~\bibnamefont {Haiman}},\ and\ \bibinfo {author} {\bibfnamefont {S.}~\bibnamefont {M\'arka}},\ }\bibfield  {title} {\bibinfo {title} {{Rapid and Bright Stellar-mass Binary Black Hole Mergers in Active Galactic Nuclei}},\ }\href {https://doi.org/10.3847/1538-4357/835/2/165} {\bibfield  {journal} {\bibinfo  {journal} {Astrophys. J.}\ }\textbf {\bibinfo {volume} {835}},\ \bibinfo {pages} {165} (\bibinfo {year} {2017})},\ \Eprint {https://arxiv.org/abs/1602.03831} {arXiv:1602.03831 [astro-ph.HE]} \BibitemShut {NoStop}%
\bibitem [{\citenamefont {Schumacher}\ \emph {et~al.}(2023)\citenamefont {Schumacher}, \citenamefont {Yunes},\ and\ \citenamefont {Yagi}}]{GW_speed}%
  \BibitemOpen
  \bibfield  {author} {\bibinfo {author} {\bibfnamefont {K.}~\bibnamefont {Schumacher}}, \bibinfo {author} {\bibfnamefont {N.}~\bibnamefont {Yunes}},\ and\ \bibinfo {author} {\bibfnamefont {K.}~\bibnamefont {Yagi}},\ }\bibfield  {title} {\bibinfo {title} {{Gravitational wave polarizations with different propagation speeds}},\ }\href {https://doi.org/10.1103/PhysRevD.108.104038} {\bibfield  {journal} {\bibinfo  {journal} {Phys. Rev. D}\ }\textbf {\bibinfo {volume} {108}},\ \bibinfo {pages} {104038} (\bibinfo {year} {2023})},\ \Eprint {https://arxiv.org/abs/2308.05589} {arXiv:2308.05589 [gr-qc]} \BibitemShut {NoStop}%
\bibitem [{\citenamefont {Tinto}\ and\ \citenamefont {Dhurandhar}(2021)}]{TDI}%
  \BibitemOpen
  \bibfield  {author} {\bibinfo {author} {\bibfnamefont {M.}~\bibnamefont {Tinto}}\ and\ \bibinfo {author} {\bibfnamefont {S.~V.}\ \bibnamefont {Dhurandhar}},\ }\bibfield  {title} {\bibinfo {title} {{Time-delay interferometry}},\ }\href {https://doi.org/10.1007/s41114-020-00029-6} {\bibfield  {journal} {\bibinfo  {journal} {Living Rev. Rel.}\ }\textbf {\bibinfo {volume} {24}},\ \bibinfo {pages} {1} (\bibinfo {year} {2021})}\BibitemShut {NoStop}%
\bibitem [{\citenamefont {Takeda}\ \emph {et~al.}(2021)\citenamefont {Takeda}, \citenamefont {Morisaki},\ and\ \citenamefont {Nishizawa}}]{non_GR_event1}%
  \BibitemOpen
  \bibfield  {author} {\bibinfo {author} {\bibfnamefont {H.}~\bibnamefont {Takeda}}, \bibinfo {author} {\bibfnamefont {S.}~\bibnamefont {Morisaki}},\ and\ \bibinfo {author} {\bibfnamefont {A.}~\bibnamefont {Nishizawa}},\ }\bibfield  {title} {\bibinfo {title} {{Pure polarization test of GW170814 and GW170817 using waveforms consistent with modified theories of gravity}},\ }\href {https://doi.org/10.1103/PhysRevD.103.064037} {\bibfield  {journal} {\bibinfo  {journal} {Phys. Rev. D}\ }\textbf {\bibinfo {volume} {103}},\ \bibinfo {pages} {064037} (\bibinfo {year} {2021})},\ \Eprint {https://arxiv.org/abs/2010.14538} {arXiv:2010.14538 [gr-qc]} \BibitemShut {NoStop}%
\bibitem [{\citenamefont {Takeda}\ \emph {et~al.}(2022)\citenamefont {Takeda}, \citenamefont {Morisaki},\ and\ \citenamefont {Nishizawa}}]{non_GR_event2}%
  \BibitemOpen
  \bibfield  {author} {\bibinfo {author} {\bibfnamefont {H.}~\bibnamefont {Takeda}}, \bibinfo {author} {\bibfnamefont {S.}~\bibnamefont {Morisaki}},\ and\ \bibinfo {author} {\bibfnamefont {A.}~\bibnamefont {Nishizawa}},\ }\bibfield  {title} {\bibinfo {title} {{Search for scalar-tensor mixed polarization modes of gravitational waves}},\ }\href {https://doi.org/10.1103/PhysRevD.105.084019} {\bibfield  {journal} {\bibinfo  {journal} {Phys. Rev. D}\ }\textbf {\bibinfo {volume} {105}},\ \bibinfo {pages} {084019} (\bibinfo {year} {2022})},\ \Eprint {https://arxiv.org/abs/2105.00253} {arXiv:2105.00253 [gr-qc]} \BibitemShut {NoStop}%
\bibitem [{\citenamefont {Takeda}\ \emph {et~al.}(2024)\citenamefont {Takeda}, \citenamefont {Tsujikawa},\ and\ \citenamefont {Nishizawa}}]{non_GR_event3}%
  \BibitemOpen
  \bibfield  {author} {\bibinfo {author} {\bibfnamefont {H.}~\bibnamefont {Takeda}}, \bibinfo {author} {\bibfnamefont {S.}~\bibnamefont {Tsujikawa}},\ and\ \bibinfo {author} {\bibfnamefont {A.}~\bibnamefont {Nishizawa}},\ }\bibfield  {title} {\bibinfo {title} {{Gravitational-wave constraints on scalar-tensor gravity from a neutron star and black-hole binary GW200115}},\ }\href {https://doi.org/10.1103/PhysRevD.109.104072} {\bibfield  {journal} {\bibinfo  {journal} {Phys. Rev. D}\ }\textbf {\bibinfo {volume} {109}},\ \bibinfo {pages} {104072} (\bibinfo {year} {2024})},\ \Eprint {https://arxiv.org/abs/2311.09281} {arXiv:2311.09281 [gr-qc]} \BibitemShut {NoStop}%
\bibitem [{\citenamefont {Battista}\ \emph {et~al.}(2023)\citenamefont {Battista}, \citenamefont {De~Falco},\ and\ \citenamefont {Usseglio}}]{Einstein_Cartan1}%
  \BibitemOpen
  \bibfield  {author} {\bibinfo {author} {\bibfnamefont {E.}~\bibnamefont {Battista}}, \bibinfo {author} {\bibfnamefont {V.}~\bibnamefont {De~Falco}},\ and\ \bibinfo {author} {\bibfnamefont {D.}~\bibnamefont {Usseglio}},\ }\bibfield  {title} {\bibinfo {title} {{First post-Newtonian N-body problem in Einstein\textendash{}Cartan theory with the Weyssenhoff fluid: Lagrangian and first integrals}},\ }\href {https://doi.org/10.1140/epjc/s10052-023-11249-9} {\bibfield  {journal} {\bibinfo  {journal} {Eur. Phys. J. C}\ }\textbf {\bibinfo {volume} {83}},\ \bibinfo {pages} {112} (\bibinfo {year} {2023})},\ \Eprint {https://arxiv.org/abs/2301.08954} {arXiv:2301.08954 [gr-qc]} \BibitemShut {NoStop}%
\bibitem [{\citenamefont {De~Falco}\ and\ \citenamefont {Battista}(2023)}]{Einstein_Cartan2}%
  \BibitemOpen
  \bibfield  {author} {\bibinfo {author} {\bibfnamefont {V.}~\bibnamefont {De~Falco}}\ and\ \bibinfo {author} {\bibfnamefont {E.}~\bibnamefont {Battista}},\ }\bibfield  {title} {\bibinfo {title} {{Analytical results for binary dynamics at the first post-Newtonian order in Einstein-Cartan theory with the Weyssenhoff fluid}},\ }\href {https://doi.org/10.1103/PhysRevD.108.064032} {\bibfield  {journal} {\bibinfo  {journal} {Phys. Rev. D}\ }\textbf {\bibinfo {volume} {108}},\ \bibinfo {pages} {064032} (\bibinfo {year} {2023})},\ \Eprint {https://arxiv.org/abs/2309.00319} {arXiv:2309.00319 [gr-qc]} \BibitemShut {NoStop}%
\bibitem [{\citenamefont {De~Falco}\ \emph {et~al.}(2024)\citenamefont {De~Falco}, \citenamefont {Battista}, \citenamefont {Usseglio},\ and\ \citenamefont {Capozziello}}]{Einstein_Cartan3}%
  \BibitemOpen
  \bibfield  {author} {\bibinfo {author} {\bibfnamefont {V.}~\bibnamefont {De~Falco}}, \bibinfo {author} {\bibfnamefont {E.}~\bibnamefont {Battista}}, \bibinfo {author} {\bibfnamefont {D.}~\bibnamefont {Usseglio}},\ and\ \bibinfo {author} {\bibfnamefont {S.}~\bibnamefont {Capozziello}},\ }\bibfield  {title} {\bibinfo {title} {{Radiative losses and radiation-reaction effects at the first post-Newtonian order in Einstein\textendash{}Cartan theory}},\ }\href {https://doi.org/10.1140/epjc/s10052-024-12476-4} {\bibfield  {journal} {\bibinfo  {journal} {Eur. Phys. J. C}\ }\textbf {\bibinfo {volume} {84}},\ \bibinfo {pages} {137} (\bibinfo {year} {2024})},\ \Eprint {https://arxiv.org/abs/2401.13374} {arXiv:2401.13374 [gr-qc]} \BibitemShut {NoStop}%
\end{thebibliography}%
\end{document}